\documentclass[journal=amacgu,manuscript=article]{achemso}
\usepackage{graphicx} 
\usepackage{chemformula}
\usepackage{csquotes}
\usepackage{hyperref}
\usepackage{tabularx}
\usepackage{rotating}
\usepackage{siunitx}
\usepackage[T1]{fontenc}

\DeclareSIUnit\angstrom{\text{\AA}}
\DeclareSIUnit\bar{bar}

\newcolumntype{Y}{>{\centering\arraybackslash}X}

\author{Linus C. Erhard}
\altaffiliation{These authors contributed equally}
\email{linus.erhard@tuwien.ac.at}
\author{Johannes Sch\"{o}rghuber}
\altaffiliation{These authors contributed equally}
\author{Aleix Comas-Vives}
\author{Georg K. H. Madsen}
\affiliation{\em Institute of Materials Chemistry, TU Wien, A-1060 Vienna, Austria}
\email{georg.madsen@tuwien.ac.at}

\title{How Realistic are Idealized Copper Surfaces?
A Machine Learning Study of Rough Copper-Water Interfaces}

\begin{document}

\begin{tocentry}

\includegraphics[width=3.25in,height=1.75in,keepaspectratio]{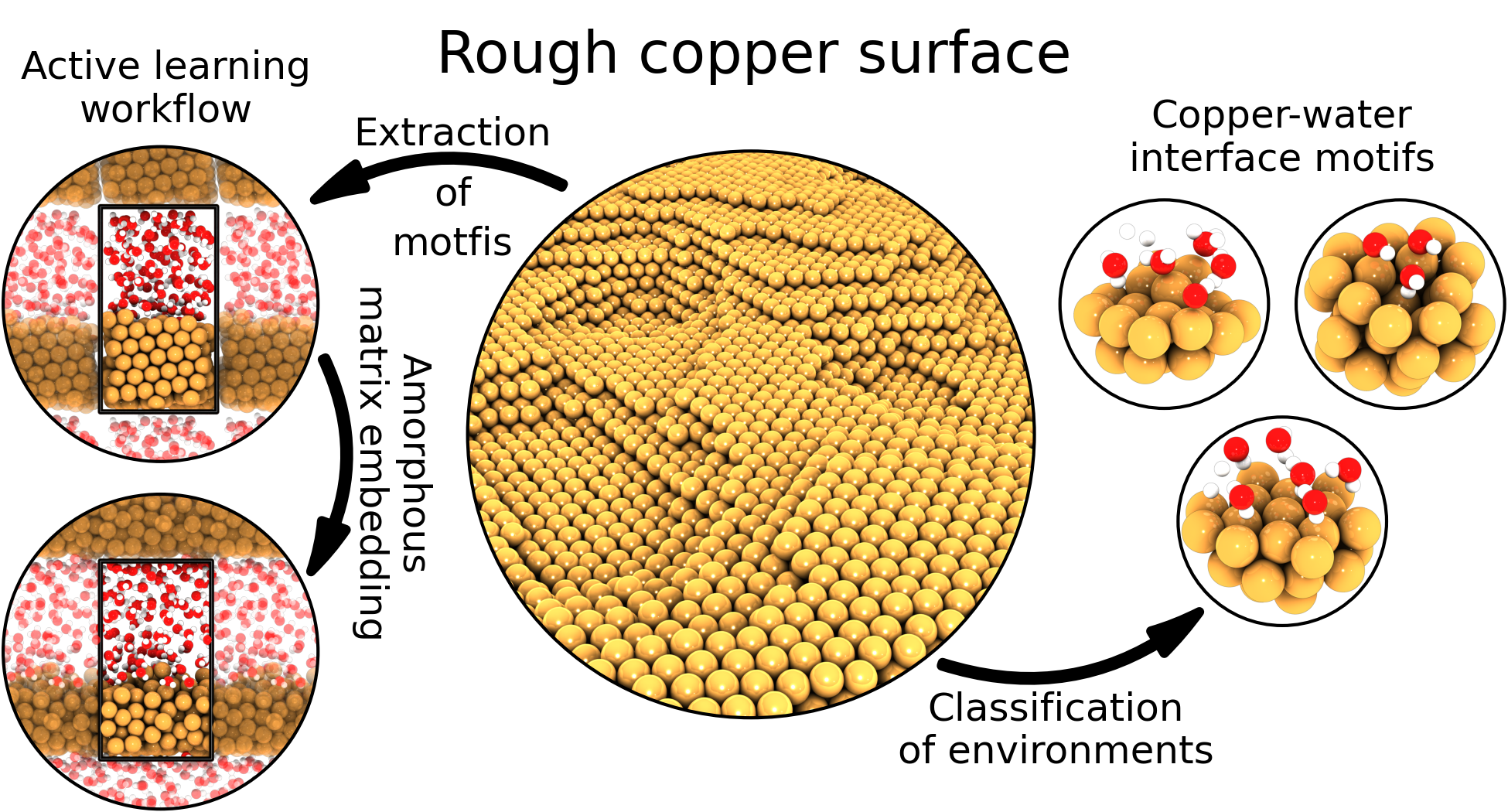}

\end{tocentry}

\begin{abstract}
    Copper is a highly promising catalyst for the electrochemical CO$_2$ reduction reaction (CO2RR) since it is the only pure metal that can form highly added-value products such as ethylene and ethanol.
    Since the CO2RR takes place in aqueous solution, the detailed atomic structure of the water-copper interface is essential for unraveling the key reaction mechanisms. 
    In this study, we investigate copper-water interfaces exhibiting nanometer-scale roughnesses. 
    We introduce two molecular dynamics protocols to create rough copper surfaces, which are subsequently brought into contact with water.
    From these interfaces, we sample additional training configurations from machine-learning-interatomic-potential-driven molecular dynamics simulations containing hundreds of thousands of atoms. An active learning workflow is developed to identify regions with high spatially resolved uncertainty and convert
them into DFT-feasible cells through a modified amorphous matrix embedding approach.
    Finally, we analyze the local environments at the interface using unsupervised machine-learning techniques. 
    Unique environments emerge on the rough copper surfaces absent from model systems, including stacking-fault-induced configurations and undercoordinated corner atoms. Notably, corner atoms consistently feature chemisorbed water molecules in our simulations, indicating their potential importance in catalytic processes.
\end{abstract}

\section*{Introduction} 

The increasing urgency of the climate crisis necessitates a rapid transition away from fossil fuels and towards renewable energy sources \cite{calvinIPCC2023Climate2023}. 
Copper is the only pure metal catalyst capable of producing substantial quantities of C2 fuels such as ethylene and ethanol via the electrochemical CO$_2$ reduction reaction.\cite{horiElectrochemicalCO2Reduction2008, nitopiProgressPerspectivesElectrochemical2019}. 
Given that
the interface between water and copper plays a critical role in the individual reaction steps, significant research efforts have been invested in the experimental characterization of these interfaces, utilizing techniques such as X-ray photo spectroscopy \cite{yamamotoSituXrayPhotoelectron2008}, X-ray diffraction \cite{kellerCompetitiveAnionWater2012}, or scanning tunneling microscopy \cite{mistryHydroxylSteppedCopper2024,leeHydroxylChainFormation2008, mehlhornLocalInvestigationFemtosecond2009}.

In addition to these experimental studies, a range of theoretical investigations have employed molecular dynamics (MD) simulations to study the copper-water interface, utilizing forces derived from density functional theory (DFT) or machine-learning interatomic potentials (MLIPs) \cite{heenenSolvationMetalWater2020, izvekovInitioMolecularDynamics2001, natarajanNeuralNetworkMolecular2016, natarajanSelfDiffusionSurfaceDefects2017, gadingRoleWaterContact2024, schorghuberFlatSteppedActive2025}. These simulations have provided valuable insights into the arrangement of water molecules at the interface, as well as the adsorption behavior and density fluctuations on low-index facets. Double peaks in water density profiles in the interface layer at pristine Cu(100) and Cu(111) confirm chemisorption of water molecules at ontop sites.\cite{natarajanNeuralNetworkMolecular2016,natarajanSelfDiffusionSurfaceDefects2017,schorghuberFlatSteppedActive2025} At stepped interfaces, the structure is dominated by the undercoordinated ridge sites, which serve as the primary sites for chemisorption.\cite{schorghuberFlatSteppedActive2025}

A significant limitation of these simulations is that they have primarily focused on ideal low-index copper surfaces, neglecting the complexity of rough surfaces, which can substantially influence the reaction mechanisms. \textit{In situ} experiments have observed the roughening of the surface, with characteristics ranging from the \AA ngstrom scale \cite{auerInterfacialWaterStructure2021} to height profiles with nanometer-scale variations \cite{auerCu111SingleCrystal2021a, simonPotentialDependentMorphologyCopper2021}, and island sizes between 5 and \SI{15}{\angstrom} \cite{auerCu111SingleCrystal2021a}. Notably, recent studies have shown that, under CO$_2$ reduction conditions, planar Cu(111) and Cu(100) surfaces are less stable than kinked surfaces \cite{chengStructureSensitivityCatalyst2025}, significantly increasing the step density under reaction conditions by a factor of 4.1 and 5.2 for the Cu(111) and Cu(100) surfaces. Furthermore, several experimental studies have observed restructuring of copper surfaces \textit{in situ} \cite{gunathungeSpectroscopicObservationReversible2017, kimEvolutionPolycrystallineCopper2014, huangPotentialinducedNanoclusteringMetallic2018, leeOxidationStateSurface2021, amirbeigiarabAtomicscaleSurfaceRestructuring2023}. Restructuring and corresponding roughening are particularly significant, as recent studies have shown that the CO$_2$ reduction reaction tends to occur on steps and kinks, rather than on planar surfaces \cite{chengStructureSensitivityCatalyst2025, nguyenInfluenceMesoscopicSurface2024}. Moreover, other studies have shown that a rougher surface significantly increases the selectivity towards the production of C2+ products \cite{jiangEffectsSurfaceRoughness2020, ebaidProductionC2C3Oxygenates2020}.

These findings emphasize the need for atomistic simulations of rough copper surfaces of nanometric size. 
However, such simulations pose a significant challenge, as they are beyond the reach of traditional DFT methods, which are limited to a few hundred atoms. 
Recent advances in MLIPs have enabled the simulation of much larger systems, with efficient feature-based techniques \cite{drautzAtomicClusterExpansion2019,thompsonSpectralNeighborAnalysis2015a} facilitating simulations of up to a billion atoms \cite{nguyen-congBillionAtomMolecular2021} or long simulations of up to tens of nanoseconds \cite{erhardUnderstandingPhaseTransitions2025}. At the same time, equivariant graph representations \cite{batznerE3equivariantGraphNeural2022, bochkarevGraphAtomicCluster2024, batatiaMACEHigherOrder2022} have substantially improved the accuracy of machine-learning interatomic potentials \cite{leimerothMachinelearningInteratomicPotentials2025}. 

Nevertheless, significant challenges persist in large-scale simulations based on MLIPs. Developing effective active learning workflows requires accurate locally resolved uncertainties that correlate directly with the true error. Recent work demonstrates that spatially averaged committee uncertainties, unlike standard committee uncertainties, provide accurate error estimation \cite{heidSpatiallyResolvedUncertainties2024}. While this enables per-atom error quantification, the necessary length scales are computationally prohibitive for direct DFT assessment. Instead, high-uncertainty environments need to be extracted into smaller representative models, which introduces boundary condition challenges. Existing mitigation strategies include embedding environments in amorphous matrices \cite{erhardModellingAtomicNanoscale2024}, minimizing boundary uncertainties \cite{lysogorskiyActiveLearningStrategies2023,kongOvercomingSizeLimit2023}, or reconstructing periodic crystalline arrangements \cite{hodappOperandoActiveLearning2020a}. 

A further distinct challenge exists in the analysis of these large-scale simulations, where surfaces are often ill-defined and environments are highly diverse. Here, unsupervised machine-learning methods are particularly valuable, as they do not rely on prior assumptions. Such techniques have previously successfully identified structural motifs on model copper-water interfaces \cite{schorghuberFlatSteppedActive2025}, detected defects in crystals \cite{kyvalaUnsupervisedIdentificationCrystal2025}, and tracked structural changes at surfaces \cite{cioniInnateDynamicsIdentity2023a}.

In this work, we leverage recent advances in MLIPs to perform atomistic simulations containing more than 200,000 atoms, allowing for a realistic representation of the diversity of rough copper surfaces. To this end, we design an active learning workflow that samples training data from the interfaces and maps it to DFT-feasible cells. This allows us to sample training data for the local environments at the rough interfaces, thereby developing suitable MLIP for copper-water interfaces that were previously out of reach. Combining this with unsupervised machine-learning techniques, we gain new insights into the structure and behavior of water at rough copper surfaces, and directly compare these findings with idealized model surfaces and the training data selected during the active learning workflow.

\section*{Computational Details}

This section describes our computational methodology, including DFT settings, MLIP fitting, and MD simulations. 
Our computational workflow relied on \texttt{ASE} \cite{hjorthlarsenAtomicSimulationEnvironment2017} and \texttt{OVITO} \cite{stukowskiVisualizationAnalysisAtomistic2010} for data processing and visualization.

\subsection*{Molecular dynamics}

We utilized the \texttt{LAMMPS} \cite{thompsonLAMMPSFlexibleSimulation2022} code for all molecular dynamics simulations in this study. We used a time step of \SI{2}{\femto\second} for all simulations including only copper, and \SI{0.5}{\femto\second} for all simulations including copper and water. We used temperature and pressure damping parameters of 100 and 1000 times the time step, respectively. Rough copper surfaces were generated with an existing Cu-Zr MLIP \cite{leimerothGeneralPurposePotential2024}, while copper-water interface simulations utilized our custom-fitted potentials described in Section \enquote{Potential fitting}.

\subsection*{Density functional theory}
\label{sec:dft}

We employed VASP 6.4.2 \cite{kresseInitioMolecularDynamics1993, kresseEfficiencyAbinitioTotal1996, kresseEfficientIterativeSchemes1996b} with the RPBE functional \cite{hammerImprovedAdsorptionEnergetics1999} for our DFT calculations. Moreover, we applied the D3 correction \cite{grimmeConsistentAccurateInitio2010} with a zero damping scheme to all atoms. 
We used an energy cutoff of \SI{850}{\eV} and Gaussian smearing with a width of \SI{0.05}{\eV}. 
The k-point density was set to correspond to a 11$\times$11$\times$11 grid for a one-atom primitive unit cell of fcc-Cu in the slab direction. 
In the direction perpendicular to the slab, a single k-point was used. We employed hard projector augmented wave pseudopotentials \cite{kresseUltrasoftPseudopotentialsProjector1999} for hydrogen and oxygen, and the standard version for copper. 
Our DFT parameters were largely based on those used in Ref.~\citenum{schorghuberFlatSteppedActive2025}. However, we applied the D3 correction to all atoms, rather than just the surface copper atoms and water molecules, due to the complexity of defining surface atoms in our rough surfaces.

\subsection*{Potential fitting}
\label{sec:potential_fitting}

We used atomic cluster expansion (ACE) and graph atomic cluster expansion (GRACE) type potentials in this study. For fitting the ACE potentials, we used the pacemaker code \cite{lysogorskiyActiveLearningStrategies2023, lysogorskiyPerformantImplementationAtomic2021, drautzAtomicClusterExpansion2019, bochkarevEfficientParametrizationAtomic2022}. Specifically, we utilized nonlinear ACE with 9 embeddings, which has been previously demonstrated to be highly effective \cite{erhardModellingAtomicNanoscale2024, leimerothGeneralPurposePotential2024}. We also explored the effects of varying the cutoff radius (\SI{5}{\angstrom},~\SI{6}{\angstrom}, and \SI{7}{\angstrom}) and the number of basis functions (900, 1200, and 1500). A detailed discussion of the parameter effects can be found in the results section. 
We used the gracemaker code \cite{bochkarevGraphAtomicCluster2024} to fit the GRACE potentials. For all GRACE models, including GRACE 1-Layer and GRACE 2-Layer, we employed the small model complexity. Additionally, for GRACE 1-Layer, we tested cutoff radii of \SI{5}{\angstrom}, \SI{6}{\angstrom}, and \SI{7}{\angstrom}. 
We used a fixed cutoff radius of \SI{5}{\angstrom} for GRACE 2-Layer. Throughout the active learning process, we consistently used a GRACE 1-Layer potential with a \SI{5}{\angstrom}~cutoff radius.

\section*{Results and Discussion}

\subsection*{Rough copper surfaces}

\begin{figure*}
    \centering
    \includegraphics[width=1.0\textwidth]{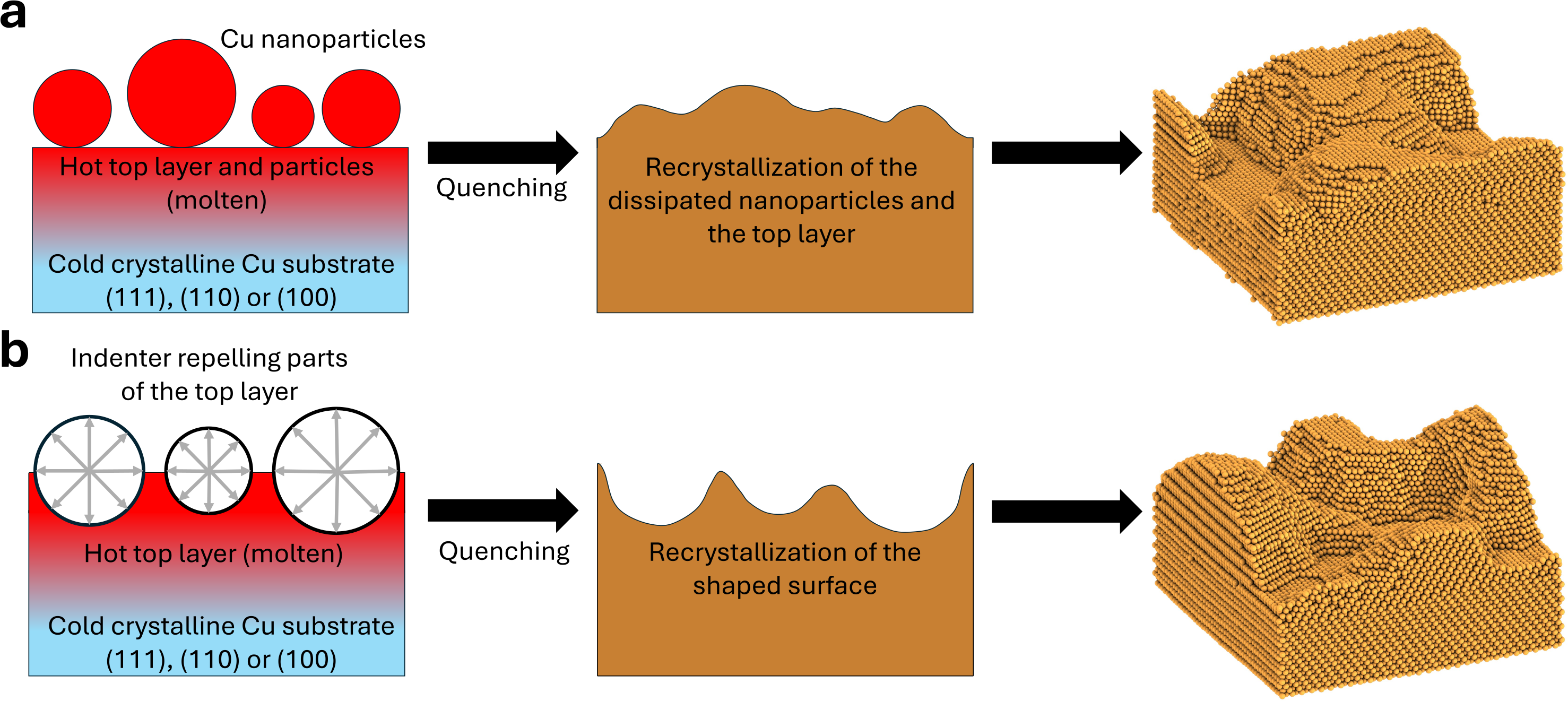}
    \caption{\textbf{Protocols to generate rough copper surfaces.} (a) In this protocol, we place copper nanoparticles on top of a copper slab. The upper part of the slab and the nanoparticles are then molten in a molecular dynamics simulation, while the lower part is kept at \SI{500}{\kelvin}. In a subsequent step, the upper part is quenched to \SI{500}{\kelvin}, initiating the recrystallization of the entire upper part. (b) In this protocol, we melt the upper part of the slab at \SI{1500}{\kelvin} while keeping the lower part at \SI{500}{\kelvin} and crystalline. Then, we press into the top layer using indenters, which push away atoms at the surface. Again, we quench the upper layer, leading to recrystallization of this part of the structure.}
    \label{fig:production_rough_copper_surface}
\end{figure*}

\autoref{fig:production_rough_copper_surface} illustrates two different methods for creating rough copper surfaces. 
In the first method, shown in \autoref{fig:production_rough_copper_surface}a, we placed copper nanoparticles on top of a copper surface slab. 
The radius of the nanoparticles was determined randomly with an average size ranging from 10 to \SI{30}{\angstrom}.
Copper slabs with Miller indices of (100), (110), and (111) were used as the substrate. 
We used supercells of 50$\times$58$\times$15 for the (111) surface, 50$\times$50$\times$17 for the (100) surface, and 35$\times$50$\times$24 for the (110) surface. 
Molecular dynamics simulations with simulations times between \SI{1}{\nano\second} and \SI{4.6}{\nano\second} (for details see Table S1) of the copper slab with nanoparticles were performed, where the upper part, including the nanoparticles, was annealed (temperatures see Table S1) while the lower part was maintained at \SI{500}{\kelvin}. 
This enabled the upper part to melt and reorganize. 
Following annealing, the upper portion was quenched to a temperature well below the melting point, inducing recrystallization. 
This approach enabled the generation of rough surfaces with relatively low surface energies and fully crystalline structures. 
Notably, the surface roughness can be controlled by varying three key factors: the initial nanoparticle size, the maximum temperature, and the quench rate. 
Two distinct MD protocols, as outlined in Table S1, were used in combination with variations in nanoparticle size to generate a range of surfaces. 
This resulted in a total of 30 surfaces, with ten surfaces generated for each of the (100), (110), and (111) slab types.

The second method, illustrated in \autoref{fig:production_rough_copper_surface}b, begins with a pristine Cu (111), (110), or (100) surface. The upper part of the surface is then heated to \SI{1500}{\kelvin}, while the lower part is maintained at \SI{500}{\kelvin}. In this top layer, indenters are inserted, which are represented by dummy particles that interact with the copper via a purely repulsive Lennard-Jones potential. These indenters repel the copper atoms at the surface, inducing roughness. The indenters are then held at a constant position, and the top layer is quenched to \SI{500}{\kelvin}, inducing recrystallization. The exact simulation protocol is outlined in Table S2. Varying the distribution of indenters, intrusion depth, and radius allows for the generation of various degrees of roughness. By randomly varying these parameters, a total of 16 surfaces were produced, comprising six (111) surfaces, and five (100) and (110) surfaces each. 

\begin{figure}
    \centering
    \includegraphics[width=0.5\linewidth]{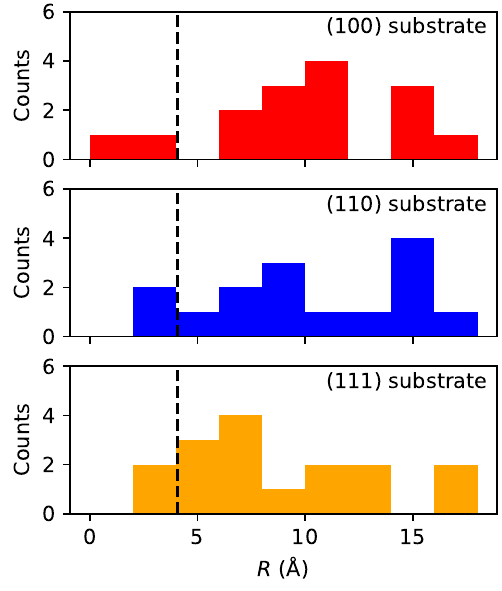}
    \caption{\textbf{Roughness of the generated copper surface for different substrates.} We show the count of the root-mean-square roughness value for all rough copper interfaces we have generated. The results are shown separately for different types of substrate surfaces. The dashed black line marks the roughness of the surface we used for our final analysis.}
    \label{fig:roughness}
\end{figure}
In \autoref{fig:roughness}, we provide the surface roughness distribution of all the 46 surfaces created. We calculate the root-mean-square roughness by,

\begin{equation}
    R = \sqrt{ \dfrac{1}{N} \sum_j^N (z_j - \overline{z})^2}, 
\end{equation}
with the number of atoms at the surface $N$, the $z_j$ coordinate of atom $j$ and the mean $z$ coordinate $\overline{z}$.
The values of the root-mean-square roughness can be as high as \SI{18}{\angstrom} and are nearly independent of the substrate surface we used. 
The smallest value of the surface roughness is around \SI{2}{\angstrom}. 
This is in good agreement with the roughest surfaces from Ref. \citenum{auerInterfacialWaterStructure2021}, which are slightly below \SI{2}{\angstrom}. 
The studies with a stronger difference in the height profile \cite{auerCu111SingleCrystal2021a,simonPotentialDependentMorphologyCopper2021} do not provide a value for the roughness. 
Since these would presumably be higher, with our structures, we are able to cover a wide range of possible surface roughness.

\subsection*{Active learning and training data}
\label{sec:rough_surfaces}

\begin{figure*}
    \centering
    \includegraphics[width=0.95\textwidth]{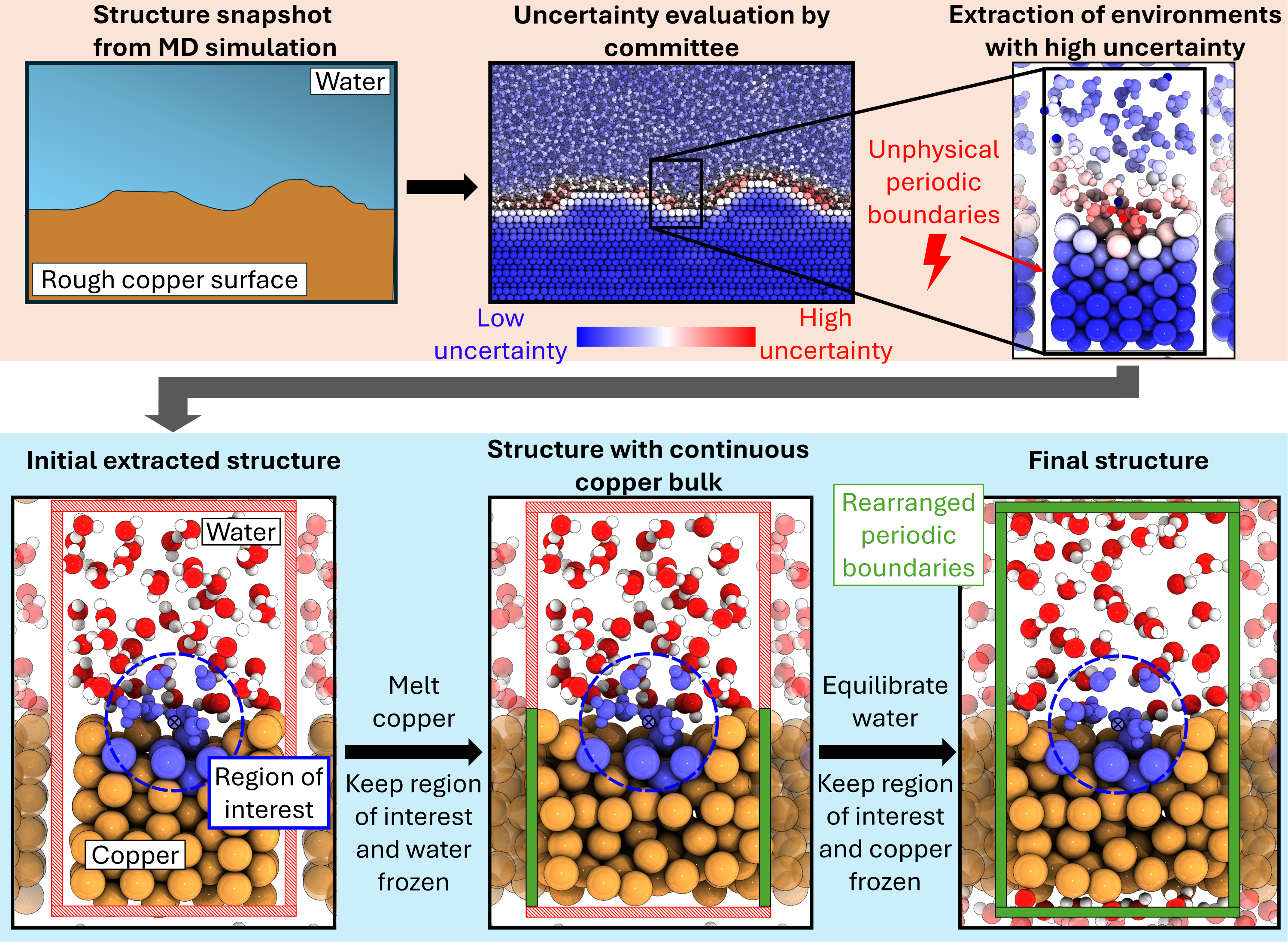}
    \caption{\textbf{Small-scale extraction of interface structures.} We sample structures from large-scale molecular dynamics simulations of rough copper-water interfaces. Using spatially resolved uncertainties based on committee errors \cite{heidSpatiallyResolvedUncertainties2024}, we determine the uncertainty for each atom. Within our simulations, the uncertainty at the interface was always the largest. Since we cannot calculate forces and energies for the entire interface, we need to extract small-scale representations. For this, we use a variation of amorphous matrix embedding \cite{erhardModellingAtomicNanoscale2024}. Within this approach, we extract a small cell from our simulation box. The atoms with the highest uncertainty are then placed centrally in the box. This central region, within a certain cutoff around the atom with high uncertainty, is the region of interest (marked blue), as training data is missing for these configurations. As we impose periodic boundary conditions in our DFT calculations, we must address the boundary artifacts that arise from inserting the extracted atoms into an arbitrary box. To fix this, we use a two-step protocol. First, we anneal the copper at high temperatures, while keeping the water and the entire area of interest fixed. This allows the copper to rearrange at the copper-copper interface, marked by the green interface area. In a second step, we equilibrate the water at room temperature, while keeping the copper and the region of interest fixed. This allows for rearrangement at the other interfaces, while the region of interest remains the same as in the beginning.}
    \label{fig:active_learning}
\end{figure*}

Accurate modelling of rough copper-water interfaces with MLIPs requires training data for these more complex interface arrangements. 
In this section, we will illustrate a way to sample training data for these rough copper-water interfaces using an active learning workflow.
As a starting point for our study, we utilized a training database for the copper-water system developed earlier in our group \cite{schorghuberFlatSteppedActive2025}. 
This database focused primarily on model surfaces like (100), (110), and (111), with additional data for more stepped surfaces like (211), (322), and (433). 
However, the database was not designed to accommodate rough copper surfaces, consisting not only of various types of edge and corner atoms, but also potentially different types of defects like vacancies or stacking faults. 

To address this limitation, we designed an active learning workflow to extract environments from large-scale simulations with high spatially resolved uncertainties \cite{heidSpatiallyResolvedUncertainties2024} and convert them into DFT-feasible cells using a modified amorphous matrix embedding approach \cite{erhardModellingAtomicNanoscale2024}. 
The workflow is illustrated in \autoref{fig:active_learning} and used as a starting point for the surfaces we generated in Fig.~ \ref{fig:production_rough_copper_surface}.

Using a committee of potentials, we evaluated the spatially resolved uncertainties within a cutoff of \SI{4}{\angstrom}~around each atom. 
We found that atoms with high uncertainty were predominantly located at the interface, in good agreement with earlier observations \cite{schorghuberFlatSteppedActive2025}. 
To calculate the forces and energy of a configuration with high uncertainty, we extracted the configuration and its surrounding area into a DFT-feasible box, as illustrated in \autoref{fig:active_learning}. 
This initial simulation box had unreasonable boundaries due to a small vacuum layer. 
To resolve this issue, we employed a two-stage annealing procedure. 
First, we annealed the copper slab to \SI{600}{\kelvin} while keeping the water and region of interest fixed. 
The region of interest was defined as all atoms within a certain cutoff around the central atom, which was the one with high uncertainty. 
This annealing step allowed us to eliminate the vacuum layer, resulting in a reasonably well-arranged, albeit amorphous, copper structure. In the second step, we repeated the process with water, letting it equilibrate at room temperature while keeping the region of interest and copper fixed. This enabled us to fill the vacuum region and re-relax the boundaries of all parts in a reasonable manner. The resulting structure was suitable for DFT convergence without significant issues, while the environment around the atom with high uncertainty remained unchanged. This approach is a modification of the recently proposed amorphous matrix embedding \cite{erhardModellingAtomicNanoscale2024}, with the key difference being the use of a two-stage annealing procedure due to the large difference in melting points between water and copper.

\begin{figure}
    \centering
    \includegraphics[width=0.5\linewidth]{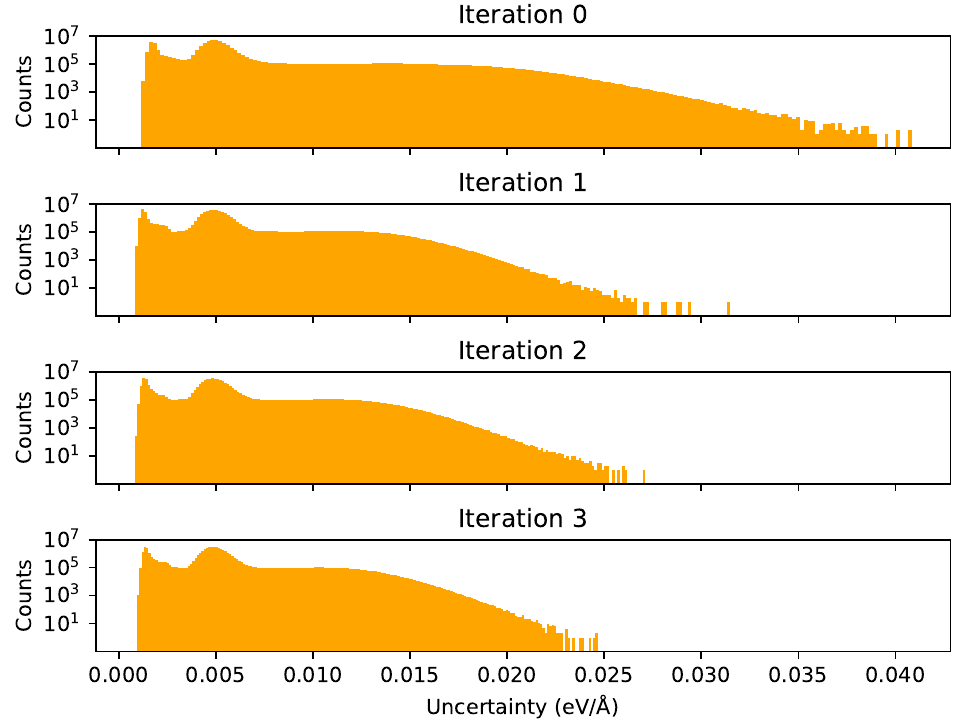}
    \caption{\textbf{Change of uncertainty over several iterations.} We evaluated the uncertainty of the committees from several iterations on the same snapshots from the last iteration's molecular dynamics run. }
    \label{fig:uncertainty_iterations}
\end{figure}
We performed a total of three active learning iterations. In the first iteration, we conducted large-scale molecular dynamics simulations on a single rough copper surface exhibiting a root-mean-square roughness of 4~\AA\ for an extended period. We then sampled 129 local environments using an uncertainty threshold of \SI{0.032}{\eV.\angstrom^{-1}}. \autoref{fig:uncertainty_iterations} shows the reduction in uncertainty of the final trajectory over AL iterations. The data generated in the first batch significantly reduced the uncertainty of the potentials. In the subsequent iterations, we sampled training data for shorter time scales for all 46 rough interfaces generated using uncertainty thresholds of \SI{0.023}{\eV.\angstrom^{-1}}~in the second iteration, and \SI{0.022}{\eV.\angstrom^{-1}}~in the third iteration to determine which configurations to add to the database. As can be seen in \autoref{fig:uncertainty_iterations}, later generations resulted in smaller improvements than the first. Using this approach, we sampled in total 238 configuration, with 129 configurations sampled in the first iteration, 44 in the second iteration and 65 in the third iteration. For the test set, we sampled configurations from environments with slightly lower uncertainties than our threshold, resulting in 38 additional configurations. This test set was then combined with the test set from Ref. \citenum{schorghuberFlatSteppedActive2025}.

\subsection*{Selection of the MLIP model}

Having successfully generated rough copper-water interfaces and created a comprehensive database that can describe these interfaces, we required a reliable machine learning interatomic potential (MLIP) for our final simulations. To this end, we compared the performance of GRACE 1-Layer, GRACE 2-Layer, and ACE models. Figure S1 presents a comparison of the computational efficiency and accuracy of these models, in terms of root-mean-square errors. While GRACE 2-Layer offers the highest accuracy, its limitation to systems with fewer than 10,000 atoms makes it unsuitable for our purposes, given that our systems comprise around 200,000 atoms. Therefore, we considered the other two models. As shown in Figure S1, ACE models exhibit significantly lower computational costs, albeit with slightly higher errors compared to GRACE 1-Layer models. Details on the parameter settings can be found in Section \enquote{Potential fitting}. Although the high computational efficiency of ACE models might make them an attractive choice, we evaluated the models based not only on their errors on the test set but also on their structural predictions at the copper-water interface. We investigated the model surfaces (100), (110), (111), (211), (322), and (433) and the arrangement of water at these surfaces, using the GRACE 2-Layer model as a reference, which is suitable for the system sizes required for the model systems. The results for the density curves of water as a function of distance to the surfaces are presented in Figure S2-S4, for oxygen and hydrogen separately and for all atoms together. While all models are in reasonable agreement with the GRACE 2-Layer reference, except the ACE potential with a \SI{5}{\angstrom}~cutoff, a detailed analysis reveals some smaller discrepancies. Notably, the first peak corresponding to chemisorbed water is underestimated for the (100) and (111) surfaces, and the hydrogen distribution deviates from the reference for the (111) surface, failing to reproduce the shape of the broad first peak. Smaller deviations are observed for the (322) and (433) surfaces, particularly in the shape of the first peak in the oxygen density function. Given that these surfaces are stepped and our interest lies in rough surfaces with many steps, these distribution functions are particularly important. Ultimately, the choice of surrogate model is challenging, but we selected the GRACE 1-Layer potential with a \SI{6}{\angstrom}~cutoff for our final simulation, as it accurately describes the interface between water and stepped Cu surfaces.

\subsection*{Interface structure} \label{sec:local_envs}

\begin{figure*}
    \includegraphics[width=\textwidth]{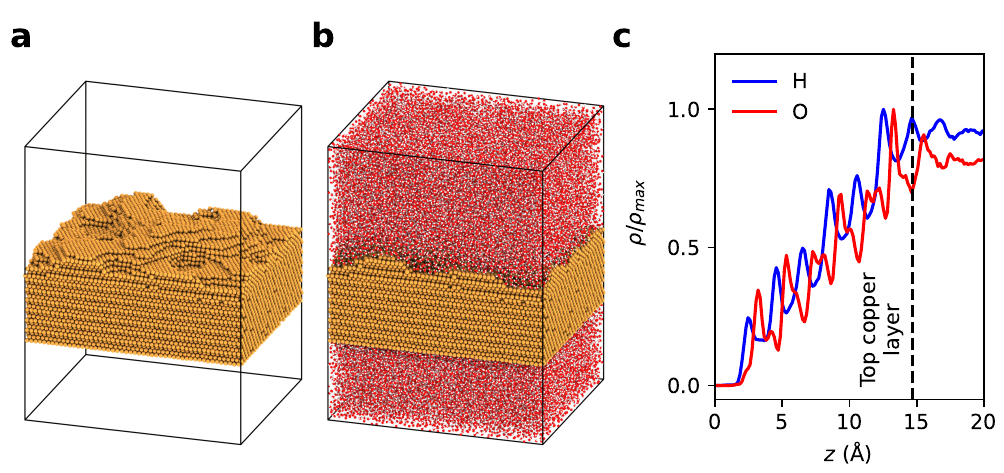}
    \caption{\textbf{Selected structure, corresponding simulation box and water distribution on the rough surface.} (a) Rough copper structure selected for detailed analysis. (b) Simulation box containing the rough copper surface and water bulk phase. (c) Hydrogen and oxygen density profiles as a function of the distance to the first surface layer, with the dotted black line indicating the highest copper layer position. The curves were normalized with respect to their global maximum.}
    \label{fig:structure_z_rdf}
\end{figure*}

\autoref{fig:structure_z_rdf}a shows the rough copper surface selected for detailed analysis of the interface. The structure was generated via the indenter method and exhibits a root-mean-square roughness of \SI{4}{\angstrom}, which represents a balanced case with significant roughness consistent with experimental measurements\cite{auerInterfacialWaterStructure2021,auerCu111SingleCrystal2021a,simonPotentialDependentMorphologyCopper2021} while avoiding extreme values. \autoref{fig:structure_z_rdf}b shows the full simulation box containing both the copper surface and water molecules. In \autoref{fig:structure_z_rdf}c hydrogen and oxygen density profiles obtained from the MD simulations of Cu-H$_2$O interfaces are shown. Due to the surface roughness, we observe increased water densities between the lowest and highest copper surface layers. The first oxygen peak appears at approximately 2.9~\AA, in agreement with earlier work on model surfaces.\cite{natarajanNeuralNetworkMolecular2016,schorghuberFlatSteppedActive2025} This is followed by periodic peaks arising from the copper layers between the lowest and highest copper surface layers. These oxygen peaks coincide with hydrogen peaks at slightly lower heights, consistent with the density profiles observed for the model stepped surfaces in the previous work and can be attributed to the water chemisorbed at step edges inducing a H-down orientation of the adjacent water molecules in order to facilitate hydrogen bonding.\cite{schorghuberFlatSteppedActive2025}

Global descriptors such as water density profiles are commonly employed to characterize the structure of metal-water interfaces in model systems defined by Miller indices. 
However, on rough surfaces, the absence of a well-defined surface onset complicates interpretation, motivating the use of alternative methods.
To capture the complexity of the interface structure, we instead employ unsupervised learning to classify local atomic environments. Similar approaches have been used to study copper-surface dynamics at elevated temperatures \cite{cioniInnateDynamicsIdentity2023a} and identify crystal defects \cite{kyvalaUnsupervisedIdentificationCrystal2025}, but we focus specifically on the interfacial copper atoms, following the methodology outlined in Ref.~\citenum{schorghuberFlatSteppedActive2025}. Local environments are encoded invariantly using spherical Bessel descriptors \cite{kocerContinuousOptimallyComplete2020,montes-camposDifferentiableNeuralNetworkForce2022} and embedded via UMAP\cite{mcinnesUMAPUniformManifold2018} for both model and the rough interface.

\begin{figure*}
    \includegraphics[width=0.95\textwidth]{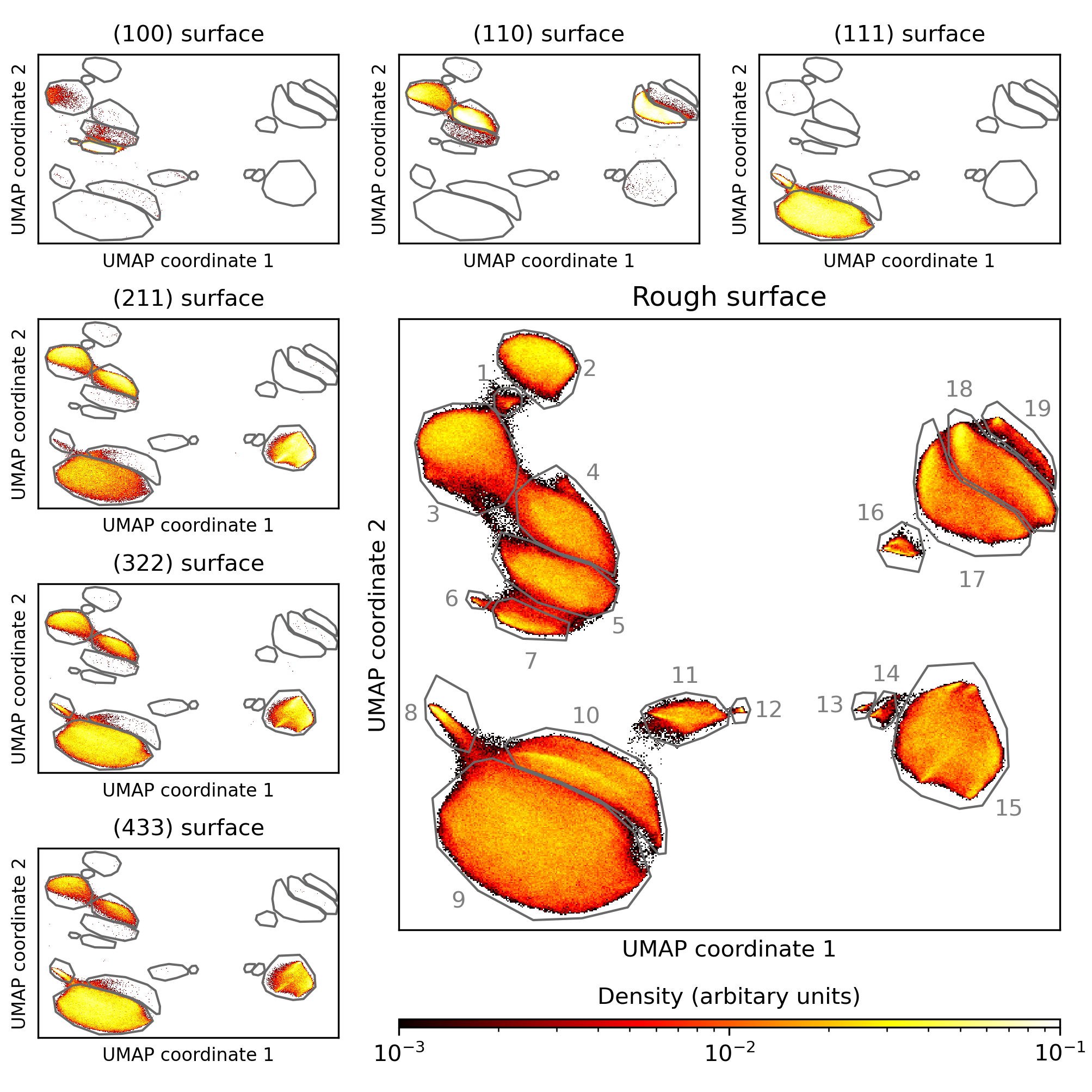}
    \caption{\textbf{UMAP embeddings of copper atoms at model surfaces and the rough surface.} The smaller panels show two-dimensional UMAP embeddings of copper atoms at various copper-water model interfaces, in particular (100), (110), (111), (211), (322), and (433). Additionally, we show in the large panel the same plot for copper surface atoms in one of our rough interface models. Moreover, we marked different clusters by gray lines and a corresponding number. These clusters are also shown for the model surfaces. Some clusters only appear on the rough copper-water interface, while all clusters that are part of the model surfaces also appear in the rough interface model. The UMAP embedding was fit on all descriptors obtained for the rough surface with 30 neighbors, a minimum distance of 0.1, and the Euclidean distance metric, and subsequently applied to the descriptors for the model surfaces.}
    \label{fig:umap_all}
\end{figure*}
\autoref{fig:umap_all} visualizes and enumerates 19 distinct clusters representing diverse local environments of top-layer copper atoms in the rough surface together with the UMAP embeddings of the copper-water model interfaces. The rough interface encompasses all environment types observed on the model surfaces, while also featuring additional clusters absent from any model interfaces. 
A comprehensive classification, including representative structures, Figure S5-S11, and cluster assignments, Table S3, is provided in the SI. 

The primary distinguishing feature is the copper coordination number. Clusters 1-7 all correspond to undercoordinated copper sites, including environments reminiscent of (110) or (100) facets and some step edges of the Cu($n+1,n,n$) surfaces included in the initial database. Strongly undercoordinated copper atoms are found at step-edge intersections, forming clusters 1,2, and 6, which are not represented in the model structures. 
While slopes generated by the indenters expose some (110) and (100) facets, the majority are (111), consistent with the surface being generated on a Cu(111) substrate. These environments, clusters 8 and 9, are well represented by the model surfaces.  Clusters 10-12 and 15-18 correspond to edge/corner atoms located directly beneath step edges.
Clusters 13, 14, and 16 occur exclusively at stacking faults and are therefore also unique for our rough copper surface. Finally, cluster 19 corresponds to atoms at the bottom of surface vacancies. 
Most of these latter clusters capture defect-like environments inaccessible to idealized surface models. Our unsupervised classification approach thus uncovers structural motifs without prior system knowledge.

\begin{figure*}
    \includegraphics[width=0.95\textwidth]{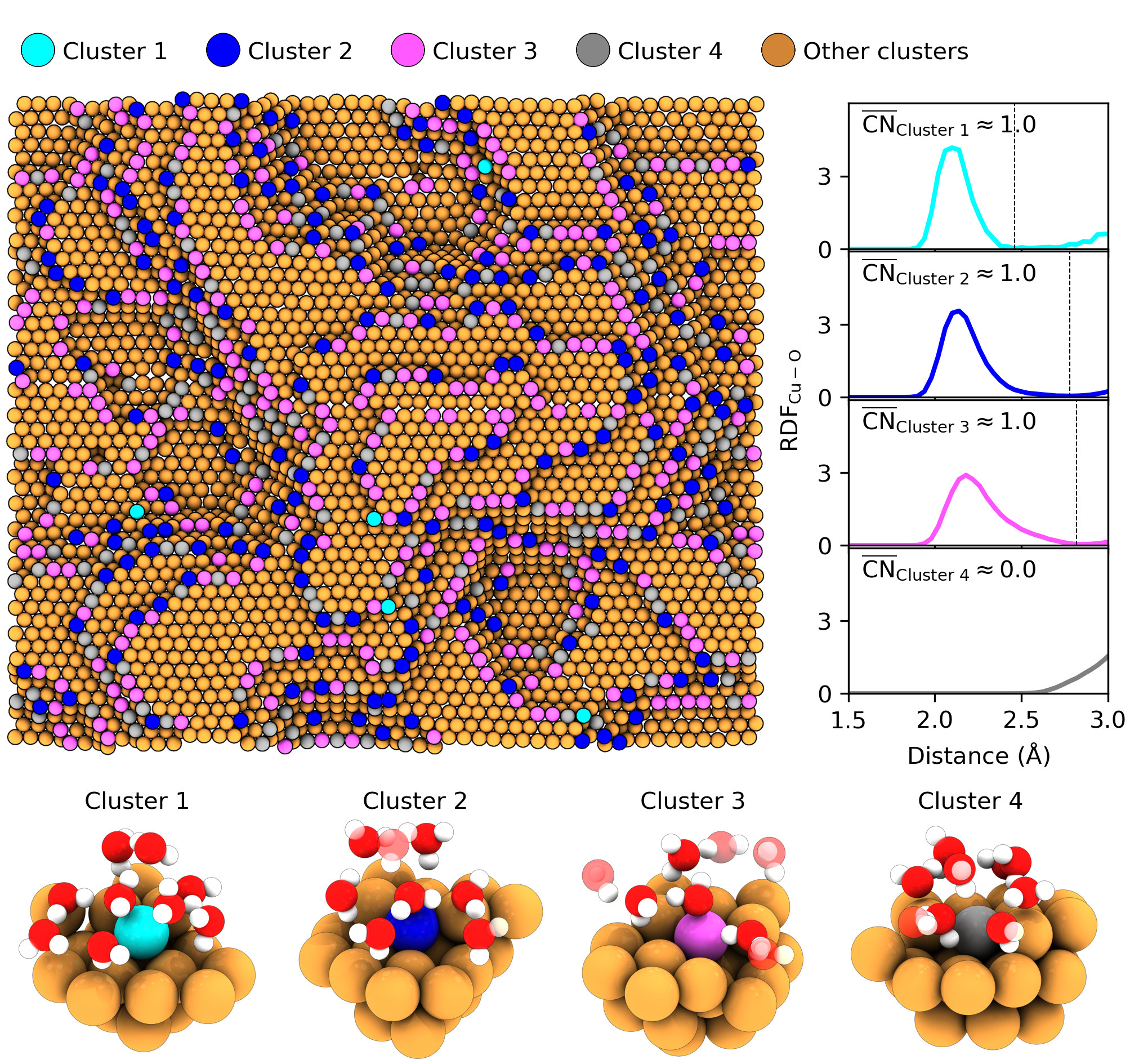}
    \caption{\textbf{Cluster classification on a rough copper surface.} Examples of clusters 1-4 from \autoref{fig:umap_all}. On the upper left we show copper atoms colored according to cluster assignment on our analyzed rough copper structure. On the right we display oxygen-copper radial distribution functions with corresponding coordination numbers, determined by integrating the first peak up to the first minimum (gray dotted line). A coordination number of 1 indicates a chemisorbed water molecule on-top of a copper atom, while 0 indicates no chemisorption. In the lower panel we show local atomic environments within \SI{5}{\angstrom} of representative atoms for each cluster. Semi-transparent oxygen and hydrogen atoms lie beyond the spatial cutoff but are included to demonstrate the absence of isolated hydrogen and oxygen atoms. }
    \label{fig:example_clusters}
\end{figure*}
In addition to the primary features related to the copper coordination, differences in Cu-H$_2$O coordination further distinguish clusters. This effect was already observed for the (111) surface where one cluster represented chemisorbed water while the other lacks chemisorption.\cite{schorghuberFlatSteppedActive2025} The same distinction can be recognized for the clusters 8 and 9, where cluster 8 represents chemisorbed water at (111) facets and cluster 9 (111) facets without chemisorbed water. Similarly, clusters 3 and 4 and 6 and 7, correspond to step edge enviroments differing by their water coordination.

\autoref{fig:example_clusters} illustrates the structural resolution achieved through our clustering approach by presenting cluster 1-4 as representative cases. 
The proximity of these clusters in the UMAP embedding (\autoref{fig:umap_all}) reflects similar local copper environments.
We will first focus on clusters 3~and~4. Both clusters represent edges of stepped (111) facets and closely resemble environments found on stepped surfaces like (211), (322), and (433), as reflected by the appearance of similar clusters in their UMAP embeddings. However, water adsorbs directly on top of copper atoms in cluster 3, while no analogous adsorption is observed in cluster 4.
This distinction was quantified through partial radial distribution functions (\autoref{fig:example_clusters} right panel), which yields average oxygen coordination numbers and confirms water chemisorption occurs on all atoms in cluster 3 but is absent in cluster 4.

Clusters 1 and 2 represent corner atoms on (111) facet with only two or three in-plane neighbors, \autoref{fig:example_clusters}. 
These distinctive configurations are exclusively found on rough surfaces, as low-index crystalline planes lack such sites.
Remarkably, persistent water chemisorption occurs on all atoms within these clusters, with no equivalent copper coordination environments lacking water chemisorption. This highlights a unique structure-adsorption correlation with potential catalytic relevance and underscores the necessity of modeling non-ideal surfaces in theoretical investigations. 
The radial distribution functions in \autoref{fig:example_clusters} further reveal a sharper first peak for cluster 1 relative to cluster 2, indicating stronger water binding to these atoms. This enhanced bonding aligns with the undercoordination of Cluster~1 sites and likely facilitates the approach of secondary water molecules within \SI{3}{\angstrom} of the adsorption site, as suggested by distinct features beyond the primary hydration shell.

\begin{figure}
    \centering
    \includegraphics[width=0.5\linewidth]{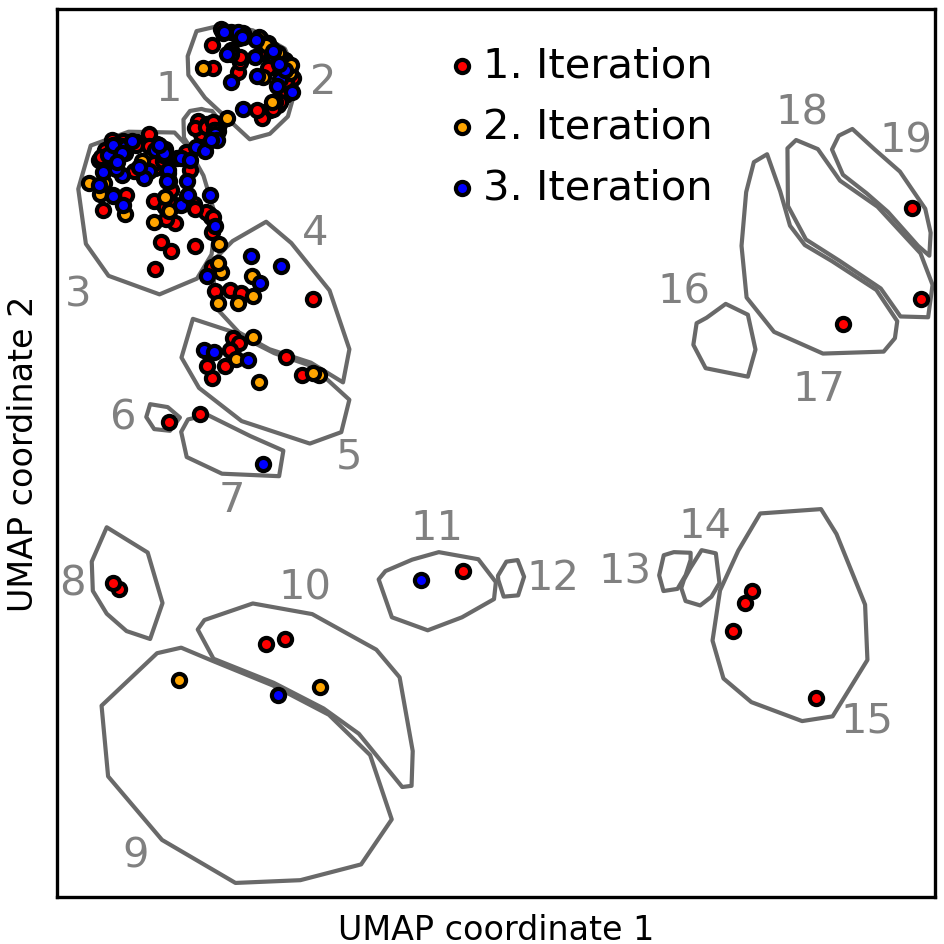}
    \caption{\textbf{Classification of environments extracted during active learning into the UMAP landscape.} The UMAP embedding shows copper atoms with the highest uncertainty from the structures extracted during active learning from the large-scale simulations. In cases where a non-copper atom had the highest uncertainty, we selected the closest copper atom to the atom with the highest uncertainty. The gray clusters are identical to those used for data classification in \autoref{fig:umap_all}.}
    \label{fig:active_learning_umap}
\end{figure}
Finally, the classification of local atomic environments provides insights into the active learning process by enabling systematic categorization of structures added during iterative sampling. Interfacial atoms consistently exhibited the highest local uncertainties, ensuring that the fixed region of interest (see workflow Fig.~\ref{fig:active_learning}) exclusively sampled these. Overall, during our active learning workflow, we sampled 238 configurations. For each configuration, we show the UMAP embedding of the central copper atom, or the closest copper atom if the central atom is not copper, in \autoref{fig:active_learning_umap}. Analysis reveals that the majority of these atoms' local environments belong to clusters 1~to~3. Clusters 1 and 2, identified above as corner sites with low in-plane copper coordination, were absent from the initial training database, confirming that spatially resolved uncertainties successfully pinpoint underrepresented structural motifs.   
Interestingly, the significant representation in cluster 3 suggests that additional training data for chemisorbed water at edge sites was required for accurate system description. This is unexpected, as these sites should be covered by prior training data from low-index copper-water interfaces, see Ref. \citenum{schorghuberFlatSteppedActive2025}. We also observe that clusters 11, 12, 13, 14, 18, and 19 are rarely sampled despite being unique to the rough interface. This rarity may arise because these environments may already be similar to bulk copper (due to minimal surface exposure) and lack water adsorption, reducing chemical complexity and subsequently the uncertainty. 
Finally, we can assess how the distribution of environments changes over the active learning cycles. The initial cycle features a broad sampling across clusters, most of which represent surfaces not previously observed by the model. However, in subsequent cycles, the sampling shows a preference for the corner-like sites within clusters 1, 2, and 3.
In summary, our workflow’s ability to automatically target complex geometries underscores a key advantage of local uncertainty quantification over structure-wide measures, which would fail to enable such precise sampling.

\section*{Conclusion} \label{sec:conclusion}

We present an explicit atomistic simulation of a rough metal-water interface. To develop an accurate machine-learning interatomic potential (MLIP) for this system, we implemented an active learning workflow that leverages spatially resolved uncertainties to identify substructures within the  interface. This approach enabled the extraction of DFT-feasible configurations and provides a generalizable workflow for generating MLIP training data for large-scale systems. 

Unsupervised classification of local atomic environments reveals a diverse ensemble of structural motifs at the Cu-H$_2$O interface. While individual model interfaces capture subsets of these environments, they fail to capture the full complexity present on rough surfaces. Water chemisorption occurs predominantly at undercoordinated Cu atoms, with step-edge intersections showing the strongest binding, suggesting these sites as primary candidates for catalytic reaction centers. In contrast, chemisorption on terrace atoms accounts for only a minor fraction of the observed sites. Our unsupervised approach identifies chemically distinct local structures without requiring a priori knowledge of relevant features. 

This work establishes a foundation for simulating interfaces with realistic surface morphologies, while also highlighting directions for future investigation. Although the present  focus is on qualitative interfacial structure assessment, extensions could include carbon-based reactants, e.g. by monitoring carbon environments to identify species such as CO$_2$ and CO, as well as variable copper oxidation states, which play an important role in CO$_2$ reduction \cite{favaroSubsurfaceOxide2017,leeOxidationStateSurface2021}. Finally, emerging methodologies \cite{bergmann2025machinelearningenergeticselectrified} may soon enable studies of electrified interfaces and their influence on structural properties.

\section*{Data availability}
The MLIP models we trained, the input files for training, the databases, and the rough copper surfaces will be made available on Zenodo upon publication under \url{https://doi.org/10.5281/zenodo.17119744}. 

\section*{Acknowledgement}
This research was funded in part by the Austrian Science Fund (FWF) 10.55776/COE5 and 10.55776/PAT3175423. The computational results presented have been achieved in part using the Vienna Scientifc Cluster (VSC). For open access purposes, the authors have applied a CC BY public copyright license to any author accepted manuscript version arising from this submission. The work leading to this publication was supported by the PRIME programme of the German Academic Exchange Service (DAAD) with funds from the German Federal Ministry of Research, Technology and Space (BMFTR).

\clearpage

\bibliography{Copper-Water.bib}

\providecommand{\latin}[1]{#1}
\makeatletter
\providecommand{\doi}
  {\begingroup\let\do\@makeother\dospecials
  \catcode`\{=1 \catcode`\}=2 \doi@aux}
\providecommand{\doi@aux}[1]{\endgroup\texttt{#1}}
\makeatother
\providecommand*\mcitethebibliography{\thebibliography}
\csname @ifundefined\endcsname{endmcitethebibliography}
  {\let\endmcitethebibliography\endthebibliography}{}
\begin{mcitethebibliography}{59}
\providecommand*\natexlab[1]{#1}
\providecommand*\mciteSetBstSublistMode[1]{}
\providecommand*\mciteSetBstMaxWidthForm[2]{}
\providecommand*\mciteBstWouldAddEndPuncttrue
  {\def\EndOfBibitem{\unskip.}}
\providecommand*\mciteBstWouldAddEndPunctfalse
  {\let\EndOfBibitem\relax}
\providecommand*\mciteSetBstMidEndSepPunct[3]{}
\providecommand*\mciteSetBstSublistLabelBeginEnd[3]{}
\providecommand*\EndOfBibitem{}
\mciteSetBstSublistMode{f}
\mciteSetBstMaxWidthForm{subitem}{(\alph{mcitesubitemcount})}
\mciteSetBstSublistLabelBeginEnd
  {\mcitemaxwidthsubitemform\space}
  {\relax}
  {\relax}

\bibitem[Calvin \latin{et~al.}(2023)Calvin, Dasgupta, Krinner, Mukherji,
  Thorne, Trisos, Romero, Aldunce, Barrett, Blanco, Cheung, Connors, Denton,
  {Diongue-Niang}, Dodman, Garschagen, Geden, Hayward, Jones, Jotzo, Krug,
  Lasco, Lee, {Masson-Delmotte}, Meinshausen, Mintenbeck, Mokssit, Otto,
  Pathak, Pirani, Poloczanska, P{\"o}rtner, Revi, Roberts, Roy, Ruane, Skea,
  Shukla, Slade, Slangen, Sokona, S{\"o}rensson, Tignor, Van~Vuuren, Wei,
  Winkler, Zhai, Zommers, Hourcade, Johnson, Pachauri, Simpson, Singh, Thomas,
  Totin, Arias, Bustamante, Elgizouli, Flato, Howden, {M{\'e}ndez-Vallejo},
  Pereira, {Pichs-Madruga}, Rose, Saheb, S{\'a}nchez~Rodr{\'i}guez,
  {\"U}rge-Vorsatz, Xiao, Yassaa, Alegr{\'i}a, Armour, {Bednar-Friedl}, Blok,
  Ciss{\'e}, Dentener, Eriksen, Fischer, Garner, Guivarch, Haasnoot, Hansen,
  Hauser, Hawkins, Hermans, Kopp, {Leprince-Ringuet}, Lewis, Ley, Ludden,
  Niamir, Nicholls, Some, Szopa, Trewin, Van Der~Wijst, Winter, Witting, Birt,
  Ha, Romero, Kim, Haites, Jung, Stavins, Birt, Ha, Orendain, Ignon, Park,
  Park, Reisinger, Cammaramo, Fischlin, Fuglestvedt, Hansen, Ludden,
  {Masson-Delmotte}, Matthews, Mintenbeck, Pirani, Poloczanska,
  {Leprince-Ringuet}, and P{\'e}an]{calvinIPCC2023Climate2023}
Calvin,~K. \latin{et~al.}  \emph{{{IPCC}}, 2023: {{Climate Change}} 2023:
  {{Synthesis Report}}. {{Contribution}} of {{Working Groups I}}, {{II}} and
  {{III}} to the {{Sixth Assessment Report}} of the {{Intergovernmental Panel}}
  on {{Climate Change}} [{{Core Writing Team}}, {{H}}. {{Lee}} and {{J}}.
  {{Romero}} (Eds.)]. {{IPCC}}, {{Geneva}}, {{Switzerland}}.}; 2023\relax
\mciteBstWouldAddEndPuncttrue
\mciteSetBstMidEndSepPunct{\mcitedefaultmidpunct}
{\mcitedefaultendpunct}{\mcitedefaultseppunct}\relax
\EndOfBibitem
\bibitem[Hori(2008)]{horiElectrochemicalCO2Reduction2008}
Hori,~Y. In \emph{Modern {{Aspects}} of {{Electrochemistry}}}; Vayenas,~C.~G.,
  White,~R.~E., {Gamboa-Aldeco},~M.~E., Eds.; Springer: New York, NY, 2008; pp
  89--189\relax
\mciteBstWouldAddEndPuncttrue
\mciteSetBstMidEndSepPunct{\mcitedefaultmidpunct}
{\mcitedefaultendpunct}{\mcitedefaultseppunct}\relax
\EndOfBibitem
\bibitem[Nitopi \latin{et~al.}(2019)Nitopi, Bertheussen, Scott, Liu, Engstfeld,
  Horch, Seger, Stephens, Chan, Hahn, N{\o}rskov, Jaramillo, and
  Chorkendorff]{nitopiProgressPerspectivesElectrochemical2019}
Nitopi,~S.; Bertheussen,~E.; Scott,~S.~B.; Liu,~X.; Engstfeld,~A.~K.;
  Horch,~S.; Seger,~B.; Stephens,~I. E.~L.; Chan,~K.; Hahn,~C.;
  N{\o}rskov,~J.~K.; Jaramillo,~T.~F.; Chorkendorff,~I. Progress and
  {{Perspectives}} of {{Electrochemical CO2 Reduction}} on {{Copper}} in
  {{Aqueous Electrolyte}}. \emph{Chemical Reviews} \textbf{2019}, \emph{119},
  7610--7672\relax
\mciteBstWouldAddEndPuncttrue
\mciteSetBstMidEndSepPunct{\mcitedefaultmidpunct}
{\mcitedefaultendpunct}{\mcitedefaultseppunct}\relax
\EndOfBibitem
\bibitem[Yamamoto \latin{et~al.}(2008)Yamamoto, Bluhm, Andersson, Ketteler,
  Ogasawara, Salmeron, and Nilsson]{yamamotoSituXrayPhotoelectron2008}
Yamamoto,~S.; Bluhm,~H.; Andersson,~K.; Ketteler,~G.; Ogasawara,~H.;
  Salmeron,~M.; Nilsson,~A. In Situ X-Ray Photoelectron Spectroscopy Studies of
  Water on Metals and Oxides at Ambient Conditions. \emph{Journal of Physics:
  Condensed Matter} \textbf{2008}, \emph{20}, 184025\relax
\mciteBstWouldAddEndPuncttrue
\mciteSetBstMidEndSepPunct{\mcitedefaultmidpunct}
{\mcitedefaultendpunct}{\mcitedefaultseppunct}\relax
\EndOfBibitem
\bibitem[Keller \latin{et~al.}(2012)Keller, Saracino, Nguyen, Huynh, and
  Broekmann]{kellerCompetitiveAnionWater2012}
Keller,~H.; Saracino,~M.; Nguyen,~H.~M.; Huynh,~T. M.~T.; Broekmann,~P.
  Competitive {{Anion}}/{{Water}} and {{Cation}}/{{Water Interactions}} at
  {{Electrified Copper}}/{{Electrolyte Interfaces Probed}} by in {{Situ X-ray
  Diffraction}}. \emph{The Journal of Physical Chemistry C} \textbf{2012},
  \emph{116}, 11068--11076\relax
\mciteBstWouldAddEndPuncttrue
\mciteSetBstMidEndSepPunct{\mcitedefaultmidpunct}
{\mcitedefaultendpunct}{\mcitedefaultseppunct}\relax
\EndOfBibitem
\bibitem[Mistry \latin{et~al.}(2024)Mistry, Snowden, Darling, and
  Hodgson]{mistryHydroxylSteppedCopper2024}
Mistry,~K.; Snowden,~H.; Darling,~G.~R.; Hodgson,~A. Hydroxyl on {{Stepped
  Copper}} and Its {{Interaction}} with {{Water}}. \emph{The Journal of
  Physical Chemistry C} \textbf{2024}, \emph{128}, 13025--13033\relax
\mciteBstWouldAddEndPuncttrue
\mciteSetBstMidEndSepPunct{\mcitedefaultmidpunct}
{\mcitedefaultendpunct}{\mcitedefaultseppunct}\relax
\EndOfBibitem
\bibitem[Lee \latin{et~al.}(2008)Lee, Sorescu, Jordan, and
  Yates]{leeHydroxylChainFormation2008}
Lee,~J.; Sorescu,~D.~C.; Jordan,~K.~D.; Yates,~J. T.~J. Hydroxyl {{Chain
  Formation}} on the {{Cu}}(110) {{Surface}}: {{Watching Water Dissociation}}.
  \emph{The Journal of Physical Chemistry C} \textbf{2008}, \emph{112},
  17672--17677\relax
\mciteBstWouldAddEndPuncttrue
\mciteSetBstMidEndSepPunct{\mcitedefaultmidpunct}
{\mcitedefaultendpunct}{\mcitedefaultseppunct}\relax
\EndOfBibitem
\bibitem[Mehlhorn \latin{et~al.}(2009)Mehlhorn, Carrasco, Michaelides, and
  Morgenstern]{mehlhornLocalInvestigationFemtosecond2009}
Mehlhorn,~M.; Carrasco,~J.; Michaelides,~A.; Morgenstern,~K. Local
  {{Investigation}} of {{Femtosecond Laser Induced Dynamics}} of {{Water
  Nanoclusters}} on {{Cu}}(111). \emph{Physical Review Letters} \textbf{2009},
  \emph{103}, 026101\relax
\mciteBstWouldAddEndPuncttrue
\mciteSetBstMidEndSepPunct{\mcitedefaultmidpunct}
{\mcitedefaultendpunct}{\mcitedefaultseppunct}\relax
\EndOfBibitem
\bibitem[Heenen \latin{et~al.}(2020)Heenen, Gauthier, Kristoffersen, Ludwig,
  and Chan]{heenenSolvationMetalWater2020}
Heenen,~H.~H.; Gauthier,~J.~A.; Kristoffersen,~H.~H.; Ludwig,~T.; Chan,~K.
  Solvation at Metal/Water Interfaces: {{An}} Ab Initio Molecular Dynamics
  Benchmark of Common Computational Approaches. \emph{The Journal of Chemical
  Physics} \textbf{2020}, \emph{152}, 144703\relax
\mciteBstWouldAddEndPuncttrue
\mciteSetBstMidEndSepPunct{\mcitedefaultmidpunct}
{\mcitedefaultendpunct}{\mcitedefaultseppunct}\relax
\EndOfBibitem
\bibitem[Izvekov \latin{et~al.}(2001)Izvekov, Mazzolo, VanOpdorp, and
  Voth]{izvekovInitioMolecularDynamics2001}
Izvekov,~S.; Mazzolo,~A.; VanOpdorp,~K.; Voth,~G.~A. Ab Initio Molecular
  Dynamics Simulation of the {{Cu}}(110)--Water Interface. \emph{The Journal of
  Chemical Physics} \textbf{2001}, \emph{114}, 3248--3257\relax
\mciteBstWouldAddEndPuncttrue
\mciteSetBstMidEndSepPunct{\mcitedefaultmidpunct}
{\mcitedefaultendpunct}{\mcitedefaultseppunct}\relax
\EndOfBibitem
\bibitem[Natarajan and Behler(2016)Natarajan, and
  Behler]{natarajanNeuralNetworkMolecular2016}
Natarajan,~S.~K.; Behler,~J. Neural Network Molecular Dynamics Simulations of
  Solid--Liquid Interfaces: Water at Low-Index Copper Surfaces. \emph{Physical
  Chemistry Chemical Physics} \textbf{2016}, \emph{18}, 28704--28725\relax
\mciteBstWouldAddEndPuncttrue
\mciteSetBstMidEndSepPunct{\mcitedefaultmidpunct}
{\mcitedefaultendpunct}{\mcitedefaultseppunct}\relax
\EndOfBibitem
\bibitem[Natarajan and Behler(2017)Natarajan, and
  Behler]{natarajanSelfDiffusionSurfaceDefects2017}
Natarajan,~S.~K.; Behler,~J. Self-{{Diffusion}} of {{Surface Defects}} at
  {{Copper}}--{{Water Interfaces}}. \emph{The Journal of Physical Chemistry C}
  \textbf{2017}, \emph{121}, 4368--4383\relax
\mciteBstWouldAddEndPuncttrue
\mciteSetBstMidEndSepPunct{\mcitedefaultmidpunct}
{\mcitedefaultendpunct}{\mcitedefaultseppunct}\relax
\EndOfBibitem
\bibitem[G{\"a}ding \latin{et~al.}(2024)G{\"a}ding, Della~Balda, Lan, Konrad,
  Iannuzzi, Mei{\ss}ner, and Tocci]{gadingRoleWaterContact2024}
G{\"a}ding,~J.; Della~Balda,~V.; Lan,~J.; Konrad,~J.; Iannuzzi,~M.;
  Mei{\ss}ner,~R.~H.; Tocci,~G. The Role of the Water Contact Layer on
  Hydration and Transport at Solid/Liquid Interfaces. \emph{Proceedings of the
  National Academy of Sciences} \textbf{2024}, \emph{121}, e2407877121\relax
\mciteBstWouldAddEndPuncttrue
\mciteSetBstMidEndSepPunct{\mcitedefaultmidpunct}
{\mcitedefaultendpunct}{\mcitedefaultseppunct}\relax
\EndOfBibitem
\bibitem[Sch{\"o}rghuber \latin{et~al.}(2025)Sch{\"o}rghuber, Bu{\v c}kov{\'a},
  Heid, and Madsen]{schorghuberFlatSteppedActive2025}
Sch{\"o}rghuber,~J.; Bu{\v c}kov{\'a},~N.; Heid,~E.; Madsen,~G. K.~H. From Flat
  to Stepped: Active Learning Frameworks for Investigating Local Structure at
  Copper--Water Interfaces. \emph{Physical Chemistry Chemical Physics}
  \textbf{2025}, \emph{27}, 9169--9177\relax
\mciteBstWouldAddEndPuncttrue
\mciteSetBstMidEndSepPunct{\mcitedefaultmidpunct}
{\mcitedefaultendpunct}{\mcitedefaultseppunct}\relax
\EndOfBibitem
\bibitem[Auer \latin{et~al.}(2021)Auer, Sarabia, Winkler, Griesser, Climent,
  Feliu, and {Kunze-Liebh{\"a}user}]{auerInterfacialWaterStructure2021}
Auer,~A.; Sarabia,~F.~J.; Winkler,~D.; Griesser,~C.; Climent,~V.; Feliu,~J.~M.;
  {Kunze-Liebh{\"a}user},~J. Interfacial {{Water Structure}} as a
  {{Descriptor}} for {{Its Electro-Reduction}} on {{Ni}}({{OH}})2-{{Modified
  Cu}}(111). \emph{ACS Catalysis} \textbf{2021}, \emph{11}, 10324--10332\relax
\mciteBstWouldAddEndPuncttrue
\mciteSetBstMidEndSepPunct{\mcitedefaultmidpunct}
{\mcitedefaultendpunct}{\mcitedefaultseppunct}\relax
\EndOfBibitem
\bibitem[Auer \latin{et~al.}(2021)Auer, Sarabia, Griesser, Climent, Feliu, and
  {Kunze-Liebh{\"a}user}]{auerCu111SingleCrystal2021a}
Auer,~A.; Sarabia,~F.~J.; Griesser,~C.; Climent,~V.; Feliu,~J.~M.;
  {Kunze-Liebh{\"a}user},~J. Cu(111) Single Crystal Electrodes: {{Modifying}}
  Interfacial Properties to Tailor Electrocatalysis. \emph{Electrochimica Acta}
  \textbf{2021}, \emph{396}, 139222\relax
\mciteBstWouldAddEndPuncttrue
\mciteSetBstMidEndSepPunct{\mcitedefaultmidpunct}
{\mcitedefaultendpunct}{\mcitedefaultseppunct}\relax
\EndOfBibitem
\bibitem[Simon \latin{et~al.}(2021)Simon, Kley, and
  Roldan~Cuenya]{simonPotentialDependentMorphologyCopper2021}
Simon,~G.~H.; Kley,~C.~S.; Roldan~Cuenya,~B. Potential-{{Dependent Morphology}}
  of {{Copper Catalysts During CO2 Electroreduction Revealed}} by {{In Situ
  Atomic Force Microscopy}}. \emph{Angewandte Chemie International Edition}
  \textbf{2021}, \emph{60}, 2561--2568\relax
\mciteBstWouldAddEndPuncttrue
\mciteSetBstMidEndSepPunct{\mcitedefaultmidpunct}
{\mcitedefaultendpunct}{\mcitedefaultseppunct}\relax
\EndOfBibitem
\bibitem[Cheng \latin{et~al.}(2025)Cheng, Nguyen, Sumaria, Wei, Zhang, Gee, Li,
  {Morales-Guio}, Heyde, Roldan~Cuenya, Alexandrova, and
  Sautet]{chengStructureSensitivityCatalyst2025}
Cheng,~D.; Nguyen,~K.-L.~C.; Sumaria,~V.; Wei,~Z.; Zhang,~Z.; Gee,~W.; Li,~Y.;
  {Morales-Guio},~C.~G.; Heyde,~M.; Roldan~Cuenya,~B.; Alexandrova,~A.~N.;
  Sautet,~P. Structure {{Sensitivity}} and {{Catalyst Restructuring}} for {{CO2
  Electro-reduction}} on {{Copper}}. \emph{Nature Communications}
  \textbf{2025}, \emph{16}, 4064\relax
\mciteBstWouldAddEndPuncttrue
\mciteSetBstMidEndSepPunct{\mcitedefaultmidpunct}
{\mcitedefaultendpunct}{\mcitedefaultseppunct}\relax
\EndOfBibitem
\bibitem[Gunathunge \latin{et~al.}(2017)Gunathunge, Li, Li, Hicks, Ovalle, and
  Waegele]{gunathungeSpectroscopicObservationReversible2017}
Gunathunge,~C.~M.; Li,~X.; Li,~J.; Hicks,~R.~P.; Ovalle,~V.~J.; Waegele,~M.~M.
  Spectroscopic {{Observation}} of {{Reversible Surface Reconstruction}} of
  {{Copper Electrodes}} under {{CO2 Reduction}}. \emph{The Journal of Physical
  Chemistry C} \textbf{2017}, \emph{121}, 12337--12344\relax
\mciteBstWouldAddEndPuncttrue
\mciteSetBstMidEndSepPunct{\mcitedefaultmidpunct}
{\mcitedefaultendpunct}{\mcitedefaultseppunct}\relax
\EndOfBibitem
\bibitem[Kim \latin{et~al.}(2014)Kim, Baricuatro, Javier, Gregoire, and
  Soriaga]{kimEvolutionPolycrystallineCopper2014}
Kim,~Y.-G.; Baricuatro,~J.~H.; Javier,~A.; Gregoire,~J.~M.; Soriaga,~M.~P. The
  {{Evolution}} of the {{Polycrystalline Copper Surface}}, {{First}} to
  {{Cu}}(111) and {{Then}} to {{Cu}}(100), at a {{Fixed CO2RR Potential}}: {{A
  Study}} by {{Operando EC-STM}}. \emph{Langmuir} \textbf{2014}, \emph{30},
  15053--15056\relax
\mciteBstWouldAddEndPuncttrue
\mciteSetBstMidEndSepPunct{\mcitedefaultmidpunct}
{\mcitedefaultendpunct}{\mcitedefaultseppunct}\relax
\EndOfBibitem
\bibitem[Huang \latin{et~al.}(2018)Huang, H{\"o}rmann, Oveisi, Loiudice,
  De~Gregorio, Andreussi, Marzari, and
  Buonsanti]{huangPotentialinducedNanoclusteringMetallic2018}
Huang,~J.; H{\"o}rmann,~N.; Oveisi,~E.; Loiudice,~A.; De~Gregorio,~G.~L.;
  Andreussi,~O.; Marzari,~N.; Buonsanti,~R. Potential-Induced Nanoclustering of
  Metallic Catalysts during Electrochemical {{CO2}} Reduction. \emph{Nature
  Communications} \textbf{2018}, \emph{9}, 3117\relax
\mciteBstWouldAddEndPuncttrue
\mciteSetBstMidEndSepPunct{\mcitedefaultmidpunct}
{\mcitedefaultendpunct}{\mcitedefaultseppunct}\relax
\EndOfBibitem
\bibitem[Lee \latin{et~al.}(2021)Lee, Lin, Farmand, Landers, Feaster,
  Avil{\'e}s~Acosta, Beeman, Ye, Yano, Mehta, Davis, Jaramillo, Hahn, and
  Drisdell]{leeOxidationStateSurface2021}
Lee,~S.~H.; Lin,~J.~C.; Farmand,~M.; Landers,~A.~T.; Feaster,~J.~T.;
  Avil{\'e}s~Acosta,~J.~E.; Beeman,~J.~W.; Ye,~Y.; Yano,~J.; Mehta,~A.;
  Davis,~R.~C.; Jaramillo,~T.~F.; Hahn,~C.; Drisdell,~W.~S. Oxidation {{State}}
  and {{Surface Reconstruction}} of {{Cu}} under {{CO2 Reduction Conditions}}
  from {{In Situ X-ray Characterization}}. \emph{Journal of the American
  Chemical Society} \textbf{2021}, \emph{143}, 588--592\relax
\mciteBstWouldAddEndPuncttrue
\mciteSetBstMidEndSepPunct{\mcitedefaultmidpunct}
{\mcitedefaultendpunct}{\mcitedefaultseppunct}\relax
\EndOfBibitem
\bibitem[Amirbeigiarab \latin{et~al.}(2023)Amirbeigiarab, Tian, Herzog, Qiu,
  Bergmann, Roldan~Cuenya, and
  Magnussen]{amirbeigiarabAtomicscaleSurfaceRestructuring2023}
Amirbeigiarab,~R.; Tian,~J.; Herzog,~A.; Qiu,~C.; Bergmann,~A.;
  Roldan~Cuenya,~B.; Magnussen,~O.~M. Atomic-Scale Surface Restructuring of
  Copper Electrodes under {{CO2}} Electroreduction Conditions. \emph{Nature
  Catalysis} \textbf{2023}, \emph{6}, 837--846\relax
\mciteBstWouldAddEndPuncttrue
\mciteSetBstMidEndSepPunct{\mcitedefaultmidpunct}
{\mcitedefaultendpunct}{\mcitedefaultseppunct}\relax
\EndOfBibitem
\bibitem[Nguyen \latin{et~al.}(2024)Nguyen, Bruce, Yoon, Navarro, Scholten,
  Landwehr, Rettenmaier, Heyde, and
  Cuenya]{nguyenInfluenceMesoscopicSurface2024}
Nguyen,~K.-L.~C.; Bruce,~J.~P.; Yoon,~A.; Navarro,~J.~J.; Scholten,~F.;
  Landwehr,~F.; Rettenmaier,~C.; Heyde,~M.; Cuenya,~B.~R. The {{Influence}} of
  {{Mesoscopic Surface Structure}} on the {{Electrocatalytic Selectivity}} of
  {{CO2 Reduction}} with {{UHV-Prepared Cu}}(111) {{Single Crystals}}.
  \emph{ACS Energy Letters} \textbf{2024}, \emph{9}, 644--652\relax
\mciteBstWouldAddEndPuncttrue
\mciteSetBstMidEndSepPunct{\mcitedefaultmidpunct}
{\mcitedefaultendpunct}{\mcitedefaultseppunct}\relax
\EndOfBibitem
\bibitem[Jiang \latin{et~al.}(2020)Jiang, Huang, Zeng, Toma, Goddard, and
  Bell]{jiangEffectsSurfaceRoughness2020}
Jiang,~K.; Huang,~Y.; Zeng,~G.; Toma,~F.~M.; Goddard,~W. A.~I.; Bell,~A.~T.
  Effects of {Surface} {Roughness} on the {Electrochemical} {Reduction} of
  {CO2} over {Cu}. \emph{ACS Energy Letters} \textbf{2020}, \emph{5},
  1206--1214, Publisher: American Chemical Society\relax
\mciteBstWouldAddEndPuncttrue
\mciteSetBstMidEndSepPunct{\mcitedefaultmidpunct}
{\mcitedefaultendpunct}{\mcitedefaultseppunct}\relax
\EndOfBibitem
\bibitem[Ebaid \latin{et~al.}(2020)Ebaid, Jiang, Zhang, Drisdell, Bell, and
  Cooper]{ebaidProductionC2C3Oxygenates2020}
Ebaid,~M.; Jiang,~K.; Zhang,~Z.; Drisdell,~W.~S.; Bell,~A.~T.; Cooper,~J.~K.
  Production of {C2}/{C3} {Oxygenates} from {Planar} {Copper}
  {Nitride}-{Derived} {Mesoporous} {Copper} via {Electrochemical} {Reduction}
  of {CO2}. \emph{Chemistry of Materials} \textbf{2020}, \emph{32}, 3304--3311,
  Publisher: American Chemical Society\relax
\mciteBstWouldAddEndPuncttrue
\mciteSetBstMidEndSepPunct{\mcitedefaultmidpunct}
{\mcitedefaultendpunct}{\mcitedefaultseppunct}\relax
\EndOfBibitem
\bibitem[Drautz(2019)]{drautzAtomicClusterExpansion2019}
Drautz,~R. Atomic Cluster Expansion for Accurate and Transferable Interatomic
  Potentials. \emph{Physical Review B} \textbf{2019}, \emph{99}, 014104\relax
\mciteBstWouldAddEndPuncttrue
\mciteSetBstMidEndSepPunct{\mcitedefaultmidpunct}
{\mcitedefaultendpunct}{\mcitedefaultseppunct}\relax
\EndOfBibitem
\bibitem[Thompson \latin{et~al.}(2015)Thompson, Swiler, Trott, Foiles, and
  Tucker]{thompsonSpectralNeighborAnalysis2015a}
Thompson,~A.~P.; Swiler,~L.~P.; Trott,~C.~R.; Foiles,~S.~M.; Tucker,~G.~J.
  Spectral Neighbor Analysis Method for Automated Generation of
  Quantum-Accurate Interatomic Potentials. \emph{Journal of Computational
  Physics} \textbf{2015}, \emph{285}, 316--330\relax
\mciteBstWouldAddEndPuncttrue
\mciteSetBstMidEndSepPunct{\mcitedefaultmidpunct}
{\mcitedefaultendpunct}{\mcitedefaultseppunct}\relax
\EndOfBibitem
\bibitem[{Nguyen-Cong} \latin{et~al.}(2021){Nguyen-Cong}, Willman, Moore,
  Belonoshko, Gayatri, Weinberg, Wood, Thompson, and
  Oleynik]{nguyen-congBillionAtomMolecular2021}
{Nguyen-Cong},~K.; Willman,~J.~T.; Moore,~S.~G.; Belonoshko,~A.~B.;
  Gayatri,~R.; Weinberg,~E.; Wood,~M.~A.; Thompson,~A.~P.; Oleynik,~I.~I.
  Billion Atom Molecular Dynamics Simulations of Carbon at Extreme Conditions
  and Experimental Time and Length Scales. Proceedings of the {{International
  Conference}} for {{High Performance Computing}}, {{Networking}}, {{Storage}}
  and {{Analysis}}. St. Louis Missouri, 2021; pp 1--12\relax
\mciteBstWouldAddEndPuncttrue
\mciteSetBstMidEndSepPunct{\mcitedefaultmidpunct}
{\mcitedefaultendpunct}{\mcitedefaultseppunct}\relax
\EndOfBibitem
\bibitem[Erhard \latin{et~al.}(2025)Erhard, Otzen, Rohrer, Prescher, and
  Albe]{erhardUnderstandingPhaseTransitions2025}
Erhard,~L.~C.; Otzen,~C.; Rohrer,~J.; Prescher,~C.; Albe,~K. Understanding
  Phase Transitions of {$\alpha$}-Quartz under Dynamic Compression Conditions
  by Machine-Learning Driven Atomistic Simulations. \emph{npj Computational
  Materials} \textbf{2025}, \emph{11}, 58\relax
\mciteBstWouldAddEndPuncttrue
\mciteSetBstMidEndSepPunct{\mcitedefaultmidpunct}
{\mcitedefaultendpunct}{\mcitedefaultseppunct}\relax
\EndOfBibitem
\bibitem[Batzner \latin{et~al.}(2022)Batzner, Musaelian, Sun, Geiger, Mailoa,
  Kornbluth, Molinari, Smidt, and
  Kozinsky]{batznerE3equivariantGraphNeural2022}
Batzner,~S.; Musaelian,~A.; Sun,~L.; Geiger,~M.; Mailoa,~J.~P.; Kornbluth,~M.;
  Molinari,~N.; Smidt,~T.~E.; Kozinsky,~B. E(3)-Equivariant Graph Neural
  Networks for Data-Efficient and Accurate Interatomic Potentials. \emph{Nature
  Communications} \textbf{2022}, \emph{13}, 2453\relax
\mciteBstWouldAddEndPuncttrue
\mciteSetBstMidEndSepPunct{\mcitedefaultmidpunct}
{\mcitedefaultendpunct}{\mcitedefaultseppunct}\relax
\EndOfBibitem
\bibitem[Bochkarev \latin{et~al.}(2024)Bochkarev, Lysogorskiy, and
  Drautz]{bochkarevGraphAtomicCluster2024}
Bochkarev,~A.; Lysogorskiy,~Y.; Drautz,~R. Graph {{Atomic Cluster Expansion}}
  for {{Semilocal Interactions}} beyond {{Equivariant Message Passing}}.
  \emph{Physical Review X} \textbf{2024}, \emph{14}, 021036\relax
\mciteBstWouldAddEndPuncttrue
\mciteSetBstMidEndSepPunct{\mcitedefaultmidpunct}
{\mcitedefaultendpunct}{\mcitedefaultseppunct}\relax
\EndOfBibitem
\bibitem[Batatia \latin{et~al.}(2022)Batatia, Kovacs, Simm, Ortner, and
  Csanyi]{batatiaMACEHigherOrder2022}
Batatia,~I.; Kovacs,~D.~P.; Simm,~G.; Ortner,~C.; Csanyi,~G. {{MACE}}: {{Higher
  Order Equivariant Message Passing Neural Networks}} for {{Fast}} and
  {{Accurate Force Fields}}. \emph{Advances in Neural Information Processing
  Systems} \textbf{2022}, \emph{35}, 11423--11436\relax
\mciteBstWouldAddEndPuncttrue
\mciteSetBstMidEndSepPunct{\mcitedefaultmidpunct}
{\mcitedefaultendpunct}{\mcitedefaultseppunct}\relax
\EndOfBibitem
\bibitem[Leimeroth \latin{et~al.}(2025)Leimeroth, Erhard, Albe, and
  Rohrer]{leimerothMachinelearningInteratomicPotentials2025}
Leimeroth,~N.; Erhard,~L.~C.; Albe,~K.; Rohrer,~J. Machine-Learning Interatomic
  Potentials from a Users Perspective: A Comparison of Accuracy, Speed and Data
  Efficiency. \emph{Modelling and Simulation in Materials Science and
  Engineering} \textbf{2025}, \emph{33}, 065012\relax
\mciteBstWouldAddEndPuncttrue
\mciteSetBstMidEndSepPunct{\mcitedefaultmidpunct}
{\mcitedefaultendpunct}{\mcitedefaultseppunct}\relax
\EndOfBibitem
\bibitem[Heid \latin{et~al.}(2024)Heid, Sch{\"o}rghuber, Wanzenb{\"o}ck, and
  Madsen]{heidSpatiallyResolvedUncertainties2024}
Heid,~E.; Sch{\"o}rghuber,~J.; Wanzenb{\"o}ck,~R.; Madsen,~G. K.~H. Spatially
  {{Resolved Uncertainties}} for {{Machine Learning Potentials}}. \emph{Journal
  of Chemical Information and Modeling} \textbf{2024}, \emph{64},
  6377--6387\relax
\mciteBstWouldAddEndPuncttrue
\mciteSetBstMidEndSepPunct{\mcitedefaultmidpunct}
{\mcitedefaultendpunct}{\mcitedefaultseppunct}\relax
\EndOfBibitem
\bibitem[Erhard \latin{et~al.}(2024)Erhard, Rohrer, Albe, and
  Deringer]{erhardModellingAtomicNanoscale2024}
Erhard,~L.~C.; Rohrer,~J.; Albe,~K.; Deringer,~V.~L. Modelling Atomic and
  Nanoscale Structure in the Silicon--Oxygen System through Active Machine
  Learning. \emph{Nature Communications} \textbf{2024}, \emph{15}, 1927\relax
\mciteBstWouldAddEndPuncttrue
\mciteSetBstMidEndSepPunct{\mcitedefaultmidpunct}
{\mcitedefaultendpunct}{\mcitedefaultseppunct}\relax
\EndOfBibitem
\bibitem[Lysogorskiy \latin{et~al.}(2023)Lysogorskiy, Bochkarev, Mrovec, and
  Drautz]{lysogorskiyActiveLearningStrategies2023}
Lysogorskiy,~Y.; Bochkarev,~A.; Mrovec,~M.; Drautz,~R. Active Learning
  Strategies for Atomic Cluster Expansion Models. \emph{Physical Review
  Materials} \textbf{2023}, \emph{7}, 043801\relax
\mciteBstWouldAddEndPuncttrue
\mciteSetBstMidEndSepPunct{\mcitedefaultmidpunct}
{\mcitedefaultendpunct}{\mcitedefaultseppunct}\relax
\EndOfBibitem
\bibitem[Kong \latin{et~al.}(2023)Kong, Li, Sun, Yang, Hao, Chen, Artrith,
  Torres, Lu, and Zhou]{kongOvercomingSizeLimit2023}
Kong,~L.; Li,~J.; Sun,~L.; Yang,~H.; Hao,~H.; Chen,~C.; Artrith,~N.; Torres,~J.
  A.~G.; Lu,~Z.; Zhou,~Y. Overcoming the {{Size Limit}} of {{First Principles
  Molecular Dynamics Simulations}} with an {{In-Distribution Substructure
  Embedding Active Learner}}. 2023\relax
\mciteBstWouldAddEndPuncttrue
\mciteSetBstMidEndSepPunct{\mcitedefaultmidpunct}
{\mcitedefaultendpunct}{\mcitedefaultseppunct}\relax
\EndOfBibitem
\bibitem[Hodapp and Shapeev(2020)Hodapp, and
  Shapeev]{hodappOperandoActiveLearning2020a}
Hodapp,~M.; Shapeev,~A. In Operando Active Learning of Interatomic Interaction
  during Large-Scale Simulations. \emph{Machine Learning: Science and
  Technology} \textbf{2020}, \emph{1}, 045005\relax
\mciteBstWouldAddEndPuncttrue
\mciteSetBstMidEndSepPunct{\mcitedefaultmidpunct}
{\mcitedefaultendpunct}{\mcitedefaultseppunct}\relax
\EndOfBibitem
\bibitem[K{\'y}vala \latin{et~al.}(2025)K{\'y}vala, {Montero de Hijes}, and
  Dellago]{kyvalaUnsupervisedIdentificationCrystal2025}
K{\'y}vala,~L.; {Montero de Hijes},~P.; Dellago,~C. Unsupervised Identification
  of Crystal Defects from Atomistic Potential Descriptors. \emph{npj
  Computational Materials} \textbf{2025}, \emph{11}, 50\relax
\mciteBstWouldAddEndPuncttrue
\mciteSetBstMidEndSepPunct{\mcitedefaultmidpunct}
{\mcitedefaultendpunct}{\mcitedefaultseppunct}\relax
\EndOfBibitem
\bibitem[Cioni \latin{et~al.}(2023)Cioni, Polino, Rapetti, Pesce, Delle~Piane,
  and Pavan]{cioniInnateDynamicsIdentity2023a}
Cioni,~M.; Polino,~D.; Rapetti,~D.; Pesce,~L.; Delle~Piane,~M.; Pavan,~G.~M.
  Innate Dynamics and Identity Crisis of a Metal Surface Unveiled by Machine
  Learning of Atomic Environments. \emph{The Journal of Chemical Physics}
  \textbf{2023}, \emph{158}, 124701\relax
\mciteBstWouldAddEndPuncttrue
\mciteSetBstMidEndSepPunct{\mcitedefaultmidpunct}
{\mcitedefaultendpunct}{\mcitedefaultseppunct}\relax
\EndOfBibitem
\bibitem[Hjorth~Larsen \latin{et~al.}(2017)Hjorth~Larsen, J{\o}rgen~Mortensen,
  Blomqvist, Castelli, Christensen, Du{\l}ak, Friis, Groves, Hammer, Hargus,
  Hermes, Jennings, Bjerre~Jensen, Kermode, Kitchin, Leonhard~Kolsbjerg, Kubal,
  Kaasbjerg, Lysgaard, Bergmann~Maronsson, Maxson, Olsen, Pastewka, Peterson,
  Rostgaard, Schi{\o}tz, Sch{\"u}tt, Strange, Thygesen, Vegge, Vilhelmsen,
  Walter, Zeng, and Jacobsen]{hjorthlarsenAtomicSimulationEnvironment2017}
Hjorth~Larsen,~A. \latin{et~al.}  The Atomic Simulation Environment---a
  {{Python}} Library for Working with Atoms. \emph{Journal of Physics:
  Condensed Matter} \textbf{2017}, \emph{29}, 273002\relax
\mciteBstWouldAddEndPuncttrue
\mciteSetBstMidEndSepPunct{\mcitedefaultmidpunct}
{\mcitedefaultendpunct}{\mcitedefaultseppunct}\relax
\EndOfBibitem
\bibitem[Stukowski(2010)]{stukowskiVisualizationAnalysisAtomistic2010}
Stukowski,~A. Visualization and Analysis of Atomistic Simulation Data with
  {{OVITO}}--the {{Open Visualization Tool}}. \emph{Modelling and Simulation in
  Materials Science and Engineering} \textbf{2010}, \emph{18}, 015012\relax
\mciteBstWouldAddEndPuncttrue
\mciteSetBstMidEndSepPunct{\mcitedefaultmidpunct}
{\mcitedefaultendpunct}{\mcitedefaultseppunct}\relax
\EndOfBibitem
\bibitem[Thompson \latin{et~al.}(2022)Thompson, Aktulga, Berger, Bolintineanu,
  Brown, Crozier, {in 't Veld}, Kohlmeyer, Moore, Nguyen, Shan, Stevens,
  Tranchida, Trott, and Plimpton]{thompsonLAMMPSFlexibleSimulation2022}
Thompson,~A.~P.; Aktulga,~H.~M.; Berger,~R.; Bolintineanu,~D.~S.; Brown,~W.~M.;
  Crozier,~P.~S.; {in 't Veld},~P.~J.; Kohlmeyer,~A.; Moore,~S.~G.;
  Nguyen,~T.~D.; Shan,~R.; Stevens,~M.~J.; Tranchida,~J.; Trott,~C.;
  Plimpton,~S.~J. {{LAMMPS}} - a Flexible Simulation Tool for Particle-Based
  Materials Modeling at the Atomic, Meso, and Continuum Scales. \emph{Computer
  Physics Communications} \textbf{2022}, \emph{271}, 108171\relax
\mciteBstWouldAddEndPuncttrue
\mciteSetBstMidEndSepPunct{\mcitedefaultmidpunct}
{\mcitedefaultendpunct}{\mcitedefaultseppunct}\relax
\EndOfBibitem
\bibitem[Leimeroth \latin{et~al.}(2024)Leimeroth, Rohrer, and
  Albe]{leimerothGeneralPurposePotential2024}
Leimeroth,~N.; Rohrer,~J.; Albe,~K. General Purpose Potential for Glassy and
  Crystalline Phases of {{Cu-Zr}} Alloys Based on the {{ACE}} Formalism.
  \emph{Physical Review Materials} \textbf{2024}, \emph{8}, 043602\relax
\mciteBstWouldAddEndPuncttrue
\mciteSetBstMidEndSepPunct{\mcitedefaultmidpunct}
{\mcitedefaultendpunct}{\mcitedefaultseppunct}\relax
\EndOfBibitem
\bibitem[Kresse and Hafner(1993)Kresse, and
  Hafner]{kresseInitioMolecularDynamics1993}
Kresse,~G.; Hafner,~J. Ab Initio Molecular Dynamics for Liquid Metals.
  \emph{Physical Review B} \textbf{1993}, \emph{47}, 558--561\relax
\mciteBstWouldAddEndPuncttrue
\mciteSetBstMidEndSepPunct{\mcitedefaultmidpunct}
{\mcitedefaultendpunct}{\mcitedefaultseppunct}\relax
\EndOfBibitem
\bibitem[Kresse and Furthm{\"u}ller(1996)Kresse, and
  Furthm{\"u}ller]{kresseEfficiencyAbinitioTotal1996}
Kresse,~G.; Furthm{\"u}ller,~J. Efficiency of Ab-Initio Total Energy
  Calculations for Metals and Semiconductors Using a Plane-Wave Basis Set.
  \emph{Computational Materials Science} \textbf{1996}, \emph{6}, 15--50\relax
\mciteBstWouldAddEndPuncttrue
\mciteSetBstMidEndSepPunct{\mcitedefaultmidpunct}
{\mcitedefaultendpunct}{\mcitedefaultseppunct}\relax
\EndOfBibitem
\bibitem[Kresse and Furthm{\"u}ller(1996)Kresse, and
  Furthm{\"u}ller]{kresseEfficientIterativeSchemes1996b}
Kresse,~G.; Furthm{\"u}ller,~J. Efficient Iterative Schemes for Ab Initio
  Total-Energy Calculations Using a Plane-Wave Basis Set. \emph{Physical Review
  B} \textbf{1996}, \emph{54}, 11169--11186\relax
\mciteBstWouldAddEndPuncttrue
\mciteSetBstMidEndSepPunct{\mcitedefaultmidpunct}
{\mcitedefaultendpunct}{\mcitedefaultseppunct}\relax
\EndOfBibitem
\bibitem[Hammer \latin{et~al.}(1999)Hammer, Hansen, and
  N{\o}rskov]{hammerImprovedAdsorptionEnergetics1999}
Hammer,~B.; Hansen,~L.~B.; N{\o}rskov,~J.~K. Improved Adsorption Energetics
  within Density-Functional Theory Using Revised {{Perdew-Burke-Ernzerhof}}
  Functionals. \emph{Physical Review B} \textbf{1999}, \emph{59},
  7413--7421\relax
\mciteBstWouldAddEndPuncttrue
\mciteSetBstMidEndSepPunct{\mcitedefaultmidpunct}
{\mcitedefaultendpunct}{\mcitedefaultseppunct}\relax
\EndOfBibitem
\bibitem[Grimme \latin{et~al.}(2010)Grimme, Antony, Ehrlich, and
  Krieg]{grimmeConsistentAccurateInitio2010}
Grimme,~S.; Antony,~J.; Ehrlich,~S.; Krieg,~H. A Consistent and Accurate Ab
  Initio Parametrization of Density Functional Dispersion Correction
  ({{DFT-D}}) for the 94 Elements {{H-Pu}}. \emph{The Journal of Chemical
  Physics} \textbf{2010}, \emph{132}, 154104\relax
\mciteBstWouldAddEndPuncttrue
\mciteSetBstMidEndSepPunct{\mcitedefaultmidpunct}
{\mcitedefaultendpunct}{\mcitedefaultseppunct}\relax
\EndOfBibitem
\bibitem[Kresse and Joubert(1999)Kresse, and
  Joubert]{kresseUltrasoftPseudopotentialsProjector1999}
Kresse,~G.; Joubert,~D. From Ultrasoft Pseudopotentials to the Projector
  Augmented-Wave Method. \emph{Physical Review B} \textbf{1999}, \emph{59},
  1758--1775\relax
\mciteBstWouldAddEndPuncttrue
\mciteSetBstMidEndSepPunct{\mcitedefaultmidpunct}
{\mcitedefaultendpunct}{\mcitedefaultseppunct}\relax
\EndOfBibitem
\bibitem[Lysogorskiy \latin{et~al.}(2021)Lysogorskiy, van~der Oord, Bochkarev,
  Menon, Rinaldi, Hammerschmidt, Mrovec, Thompson, Cs{\'a}nyi, Ortner, and
  Drautz]{lysogorskiyPerformantImplementationAtomic2021}
Lysogorskiy,~Y.; van~der Oord,~C.; Bochkarev,~A.; Menon,~S.; Rinaldi,~M.;
  Hammerschmidt,~T.; Mrovec,~M.; Thompson,~A.; Cs{\'a}nyi,~G.; Ortner,~C.;
  Drautz,~R. Performant Implementation of the Atomic Cluster Expansion
  ({{PACE}}) and Application to Copper and Silicon. \emph{npj Computational
  Materials} \textbf{2021}, \emph{7}, 97\relax
\mciteBstWouldAddEndPuncttrue
\mciteSetBstMidEndSepPunct{\mcitedefaultmidpunct}
{\mcitedefaultendpunct}{\mcitedefaultseppunct}\relax
\EndOfBibitem
\bibitem[Bochkarev \latin{et~al.}(2022)Bochkarev, Lysogorskiy, Menon, Qamar,
  Mrovec, and Drautz]{bochkarevEfficientParametrizationAtomic2022}
Bochkarev,~A.; Lysogorskiy,~Y.; Menon,~S.; Qamar,~M.; Mrovec,~M.; Drautz,~R.
  Efficient Parametrization of the Atomic Cluster Expansion. \emph{Physical
  Review Materials} \textbf{2022}, \emph{6}, 013804\relax
\mciteBstWouldAddEndPuncttrue
\mciteSetBstMidEndSepPunct{\mcitedefaultmidpunct}
{\mcitedefaultendpunct}{\mcitedefaultseppunct}\relax
\EndOfBibitem
\bibitem[Kocer \latin{et~al.}(2020)Kocer, Mason, and
  Erturk]{kocerContinuousOptimallyComplete2020}
Kocer,~E.; Mason,~J.~K.; Erturk,~H. Continuous and Optimally Complete
  Description of Chemical Environments Using {{Spherical Bessel}} Descriptors.
  \emph{AIP Advances} \textbf{2020}, \emph{10}, 015021\relax
\mciteBstWouldAddEndPuncttrue
\mciteSetBstMidEndSepPunct{\mcitedefaultmidpunct}
{\mcitedefaultendpunct}{\mcitedefaultseppunct}\relax
\EndOfBibitem
\bibitem[{Montes-Campos} \latin{et~al.}(2022){Montes-Campos}, Carrete,
  Bichelmaier, Varela, and
  Madsen]{montes-camposDifferentiableNeuralNetworkForce2022}
{Montes-Campos},~H.; Carrete,~J.; Bichelmaier,~S.; Varela,~L.~M.; Madsen,~G.
  K.~H. A {{Differentiable Neural-Network Force Field}} for {{Ionic Liquids}}.
  \emph{Journal of Chemical Information and Modeling} \textbf{2022}, \emph{62},
  88--101\relax
\mciteBstWouldAddEndPuncttrue
\mciteSetBstMidEndSepPunct{\mcitedefaultmidpunct}
{\mcitedefaultendpunct}{\mcitedefaultseppunct}\relax
\EndOfBibitem
\bibitem[McInnes \latin{et~al.}(2018)McInnes, Healy, Saul, and
  Gro{\ss}berger]{mcinnesUMAPUniformManifold2018}
McInnes,~L.; Healy,~J.; Saul,~N.; Gro{\ss}berger,~L. {{UMAP}}: {{Uniform
  Manifold Approximation}} and {{Projection}}. \emph{Journal of Open Source
  Software} \textbf{2018}, \emph{3}, 861\relax
\mciteBstWouldAddEndPuncttrue
\mciteSetBstMidEndSepPunct{\mcitedefaultmidpunct}
{\mcitedefaultendpunct}{\mcitedefaultseppunct}\relax
\EndOfBibitem
\bibitem[Favaro \latin{et~al.}(2017)Favaro, Xiao, Cheng, Goddard, Yano, and
  Crumlin]{favaroSubsurfaceOxide2017}
Favaro,~M.; Xiao,~H.; Cheng,~T.; Goddard,~W.~A.; Yano,~J.; Crumlin,~E.~J.
  Subsurface oxide plays a critical role in {CO2} activation by {Cu}(111)
  surfaces to form chemisorbed {CO2}, the first step in reduction of {CO2}.
  \emph{Proceedings of the National Academy of Sciences} \textbf{2017},
  \emph{114}, 6706--6711\relax
\mciteBstWouldAddEndPuncttrue
\mciteSetBstMidEndSepPunct{\mcitedefaultmidpunct}
{\mcitedefaultendpunct}{\mcitedefaultseppunct}\relax
\EndOfBibitem
\bibitem[Bergmann \latin{et~al.}(2025)Bergmann, Bonnet, Marzari, Reuter, and
  Hörmann]{bergmann2025machinelearningenergeticselectrified}
Bergmann,~N.; Bonnet,~N.; Marzari,~N.; Reuter,~K.; Hörmann,~N.~G. Machine
  Learning the Energetics of Electrified Solid/Liquid Interfaces. 2025;
  \url{https://arxiv.org/abs/2505.19745}\relax
\mciteBstWouldAddEndPuncttrue
\mciteSetBstMidEndSepPunct{\mcitedefaultmidpunct}
{\mcitedefaultendpunct}{\mcitedefaultseppunct}\relax
\EndOfBibitem
\end{mcitethebibliography}

\end{document}


\begin{table*}[b!]
\begin{tabularx}{\textwidth}{p{1.5cm}YYYY}
    \hline
    \hline
     Protocol & Simulation Time [\si{\pico\second}] & Temperature [\si{\kelvin}] & Temperature [\si{\kelvin}] & Box Change Allowed \\
              &                      & Top Layer & Bottom Layer & in Surface Plane \\
              \hline
    \\
     Version 1 \\
     \hline
     Step 1. & 20 & 500 & 500 & Yes   \\
     Step 2. & 200 & 500$\rightarrow$ 1100 & 500 & No \\
     Step 3. & 400 & 1100 & 500 & No \\
     Step 4. & 4000 & 1100 $\rightarrow$ 500 & 500 & No \\
     \hline
     \\
     Version 2 \\
     \hline
     Step 1. & 20 & 500 & 500 & Yes   \\
     Step 2. & 20 & 500$\rightarrow$ 1300 & 500 & No \\
     Step 3. & 10 & 1300 & 500 & No \\
     Step 4. & 20 & 1300 $\rightarrow$ 900 & 500 & No \\
     Step 5. & 1000 & 900 $\rightarrow$ 500 & 500 & No \\
     \hline
     \hline
\end{tabularx}
     \caption{\textbf{Detailed simulation protocol to produce rough copper surfaces by nanoparticles placed on top.}We used two different temperature protocols for producing rough surfaces. In version 1, we used lower temperatures of maximum 1100~K, but therefore longer simulation times. In version 2, we used higher temperatures of a maximum of 1300~K, but for much shorter times. The temperature of the bottom part of the slab was always fixed at 500~K. In an intermediated region of 10~\AA~we used starting from the second step a NVE ensemble to allow for a temperature gradient.}
     \label{stab:protocol_nanoparticles}
\end{table*}

\begin{table*}[b!]
\begin{tabularx}{\textwidth}{p{1.4cm}YYYYY}
    \hline
    \hline
     Protocol & Simulation Time & Temperature [\si{\kelvin}] & Temperature [\si{\kelvin}] & Indenters & Indenters \\
              &      [\si{\pico\second}]                & Top Layer & Bottom Layer & Active & Moving Down \\
              \hline
    \\
     \hline
     Step 1. & 10  & 500 & 500 & No & No   \\
     Step 2. & 40  & 500$\rightarrow$ 1500 & 500 & No & No \\
     Step 3. & 200 & 1500 & 500 & Yes & Yes \\
     Step 4. & 40  & 1500 & 500 & Yes & No \\
     Step 5. & 2000 & 1500 $\rightarrow$ 500 & 500 & Yes & No \\
     \hline
     \hline
\end{tabularx}
     \caption{\textbf{Detailed simulation protocol to produce rough copper surfaces by inserting indenters.} The whole simulation is performed at constant volume conditions. We start by equilibrating the system and then increasing the temperature of the top layer. Between the top layer and the bottom layer is a 10~\AA~thick region, where we do not apply a thermostat. The roughness of the surface is then induced by indenters inserted from above the slab into the top layer. The indenters are implemented by dummy particles, which interact by a purely repulsive Lennard-Jones potential with the copper. The radius of the indenters, as well as the intrusion depth, is varied randomly for different structures. We apply a reflecting wall at the bottom and well above the top of the slab to prevent atoms from escaping. After inserting, we keep the indenters at a constant position in the simulation, while we cool down the top layer.}
     \label{stab:protocol_indenter}
\end{table*}

\clearpage

\begin{figure*}
    \centering
    \includegraphics[width=1.0\linewidth]{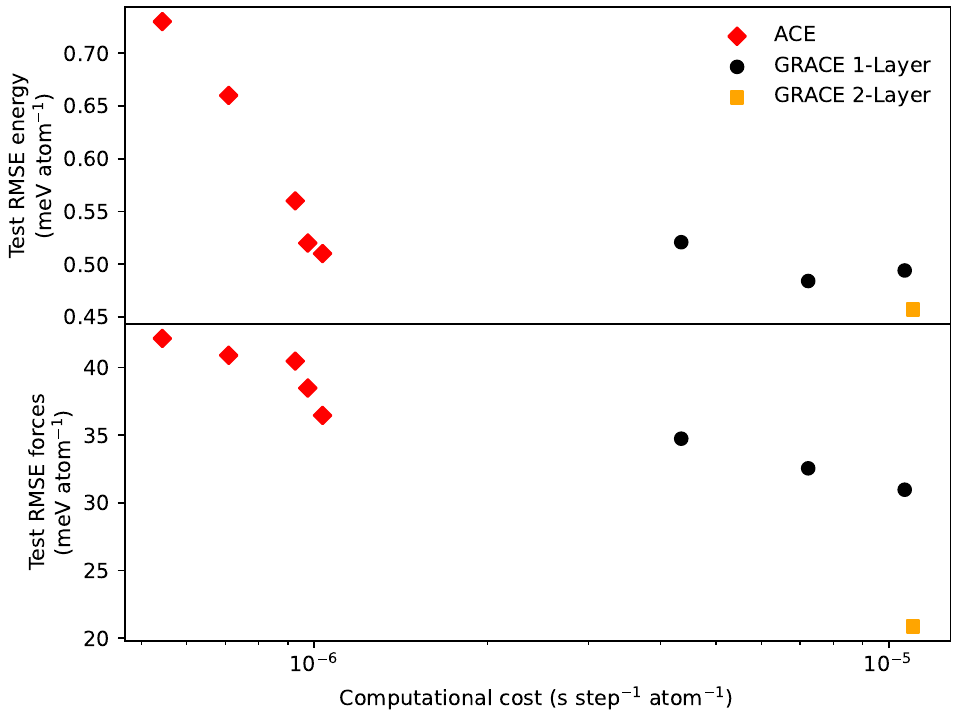}
    \caption{\textbf{Accuracy and computational costs of ACE and different types of GRACE.} ACE and GRACE 1-Layer are compared with regard to their accuracy and their computational efficiency to GRACE 2-Layer. In the case of ACE, we varied the cutoff radius and the number of basis functions, in the case of GRACE 1-Layer, we varied only the cutoff radius, leading to several points in the plot.}
    \label{sfig:pareto}
\end{figure*}

\begin{sidewaysfigure*}
    \centering
    \includegraphics[width=1.1\textwidth]{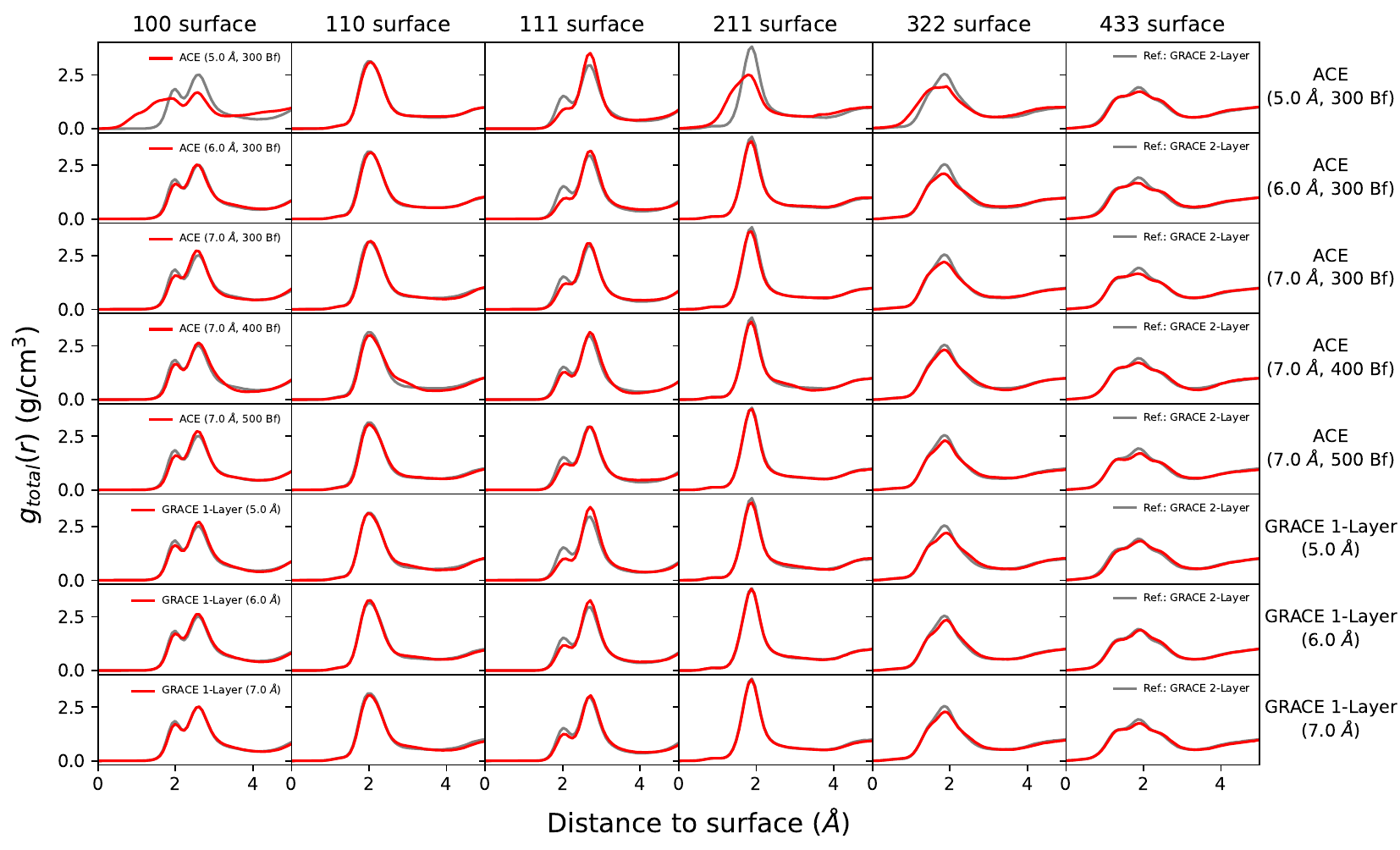}
    \caption{\textbf{Density of water as a function of the distance to various copper surfaces.} These plots show how water is distributed as a function of the distances on the copper 100, 110, 111, 211, 322, and 433 surfaces. We evaluated the curves with various machine-learning interatomic potentials, e.g., ACE, GRACE 1-Layer, and GRACE 2-Layer. All potentials have been fitted to the same databases. As GRACE 2-Layer provides the highest accuracy (see \autoref{sfig:pareto}), we used it in all plots as a reference line (gray).}
    \label{sfig:rdfs_all}
\end{sidewaysfigure*}

\begin{sidewaysfigure*}
    \centering
    \includegraphics[width=1.1\textwidth]{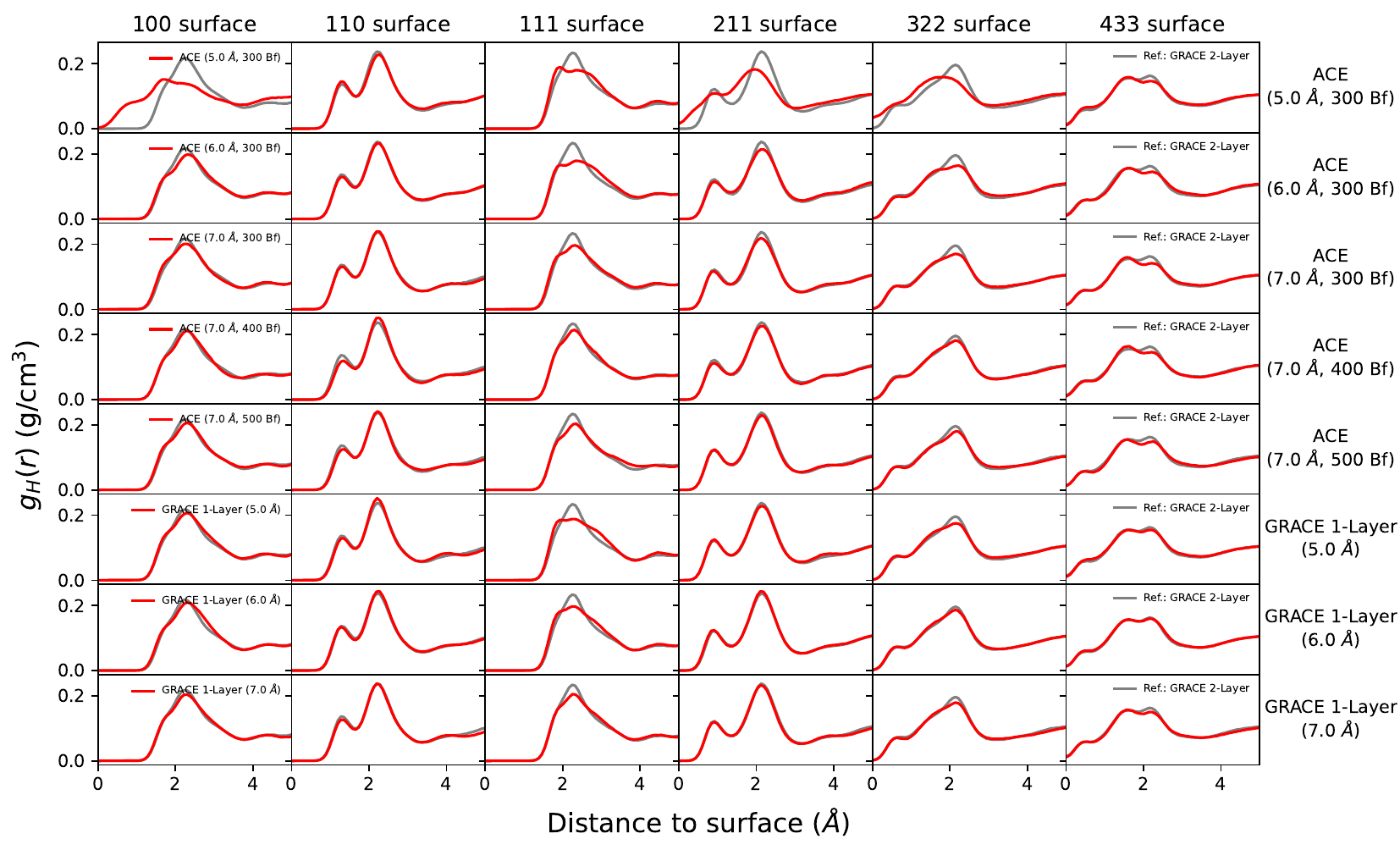}
    \caption{\textbf{Density of hydrogen in water as a function of the distance to various copper surfaces.} These plots show how hydrogen, as part of the water molecules, is distributed as a function of the distances on the copper 100, 110, 111, 211, 322, and 433 surfaces. We evaluated the curves with various machine-learning interatomic potentials, e.g., ACE, GRACE 1-Layer, and GRACE 2-Layer. All potentials have been fitted to the same databases. As GRACE 2-Layer provides the highest accuracy (see \autoref{sfig:pareto}), we used it in all plots as a reference line (gray).}
    \label{sfig:rdfs_hydrogen}
\end{sidewaysfigure*}

\begin{sidewaysfigure*}
    \centering
    \includegraphics[width=1.1\textwidth]{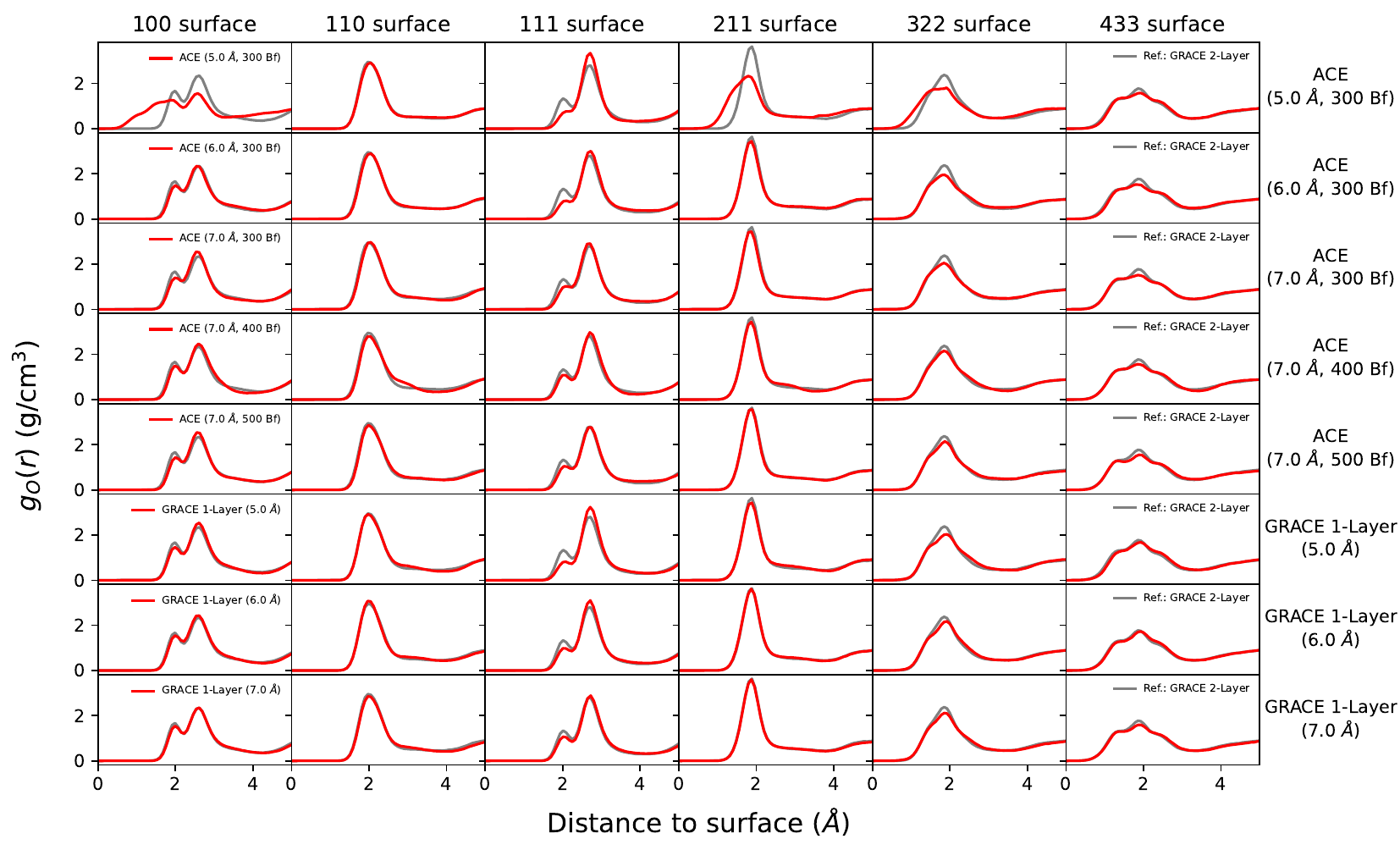}
    \caption{\textbf{Density of oxygen in water as a function of the distance to various copper surfaces.} These plots show how oxygen, as part of water molecules, is distributed as a function of the distances on the copper 100, 110, 111, 211, 322, and 433 surfaces. We evaluated the curves with various machine-learning interatomic potentials, e.g., ACE, GRACE 1-Layer, and GRACE 2-Layer. All potentials have been fitted to the same databases. As GRACE 2-Layer provides the highest accuracy (see \autoref{sfig:pareto}), we used it in all plots as a reference line (gray).}
    \label{sfig:rdfs_oxygen}
\end{sidewaysfigure*}

\begin{figure*}
    \centering
    \includegraphics[width=0.8\textwidth]{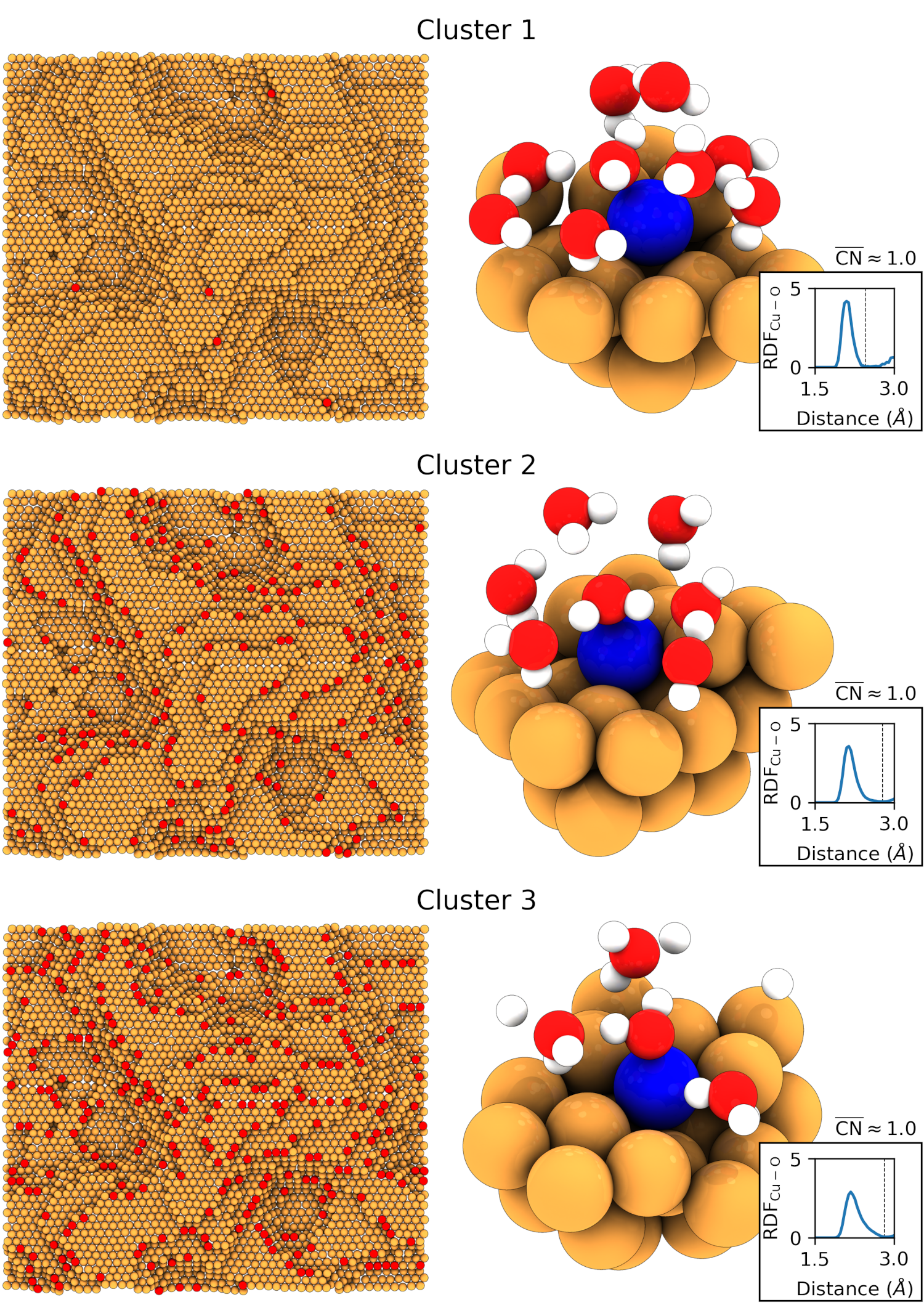}
    \caption{\textbf{Rough copper surface with atoms marked according to their cluster classification.} The left side of the figure depicts the rough copper surface, with atoms colored according to their cluster assignment as specified in Figure 6. On the right, an exemplary environment within a \SI{5}{\angstrom} radius of a cluster atom (marked in blue) is shown. In this representation, copper atoms are colored brown, oxygen atoms are colored red, and hydrogen atoms are colored white. Additionally, the average radial distribution function between copper atoms in this cluster and oxygen atoms is plotted. The average coordination number of copper with respect to oxygen is determined by integrating the first peak of the radial distribution function up to the point indicated by the black line.}
    \label{sfig:interface_clusters1}
\end{figure*}

\begin{figure*}
    \centering
    \includegraphics[width=0.8\textwidth]{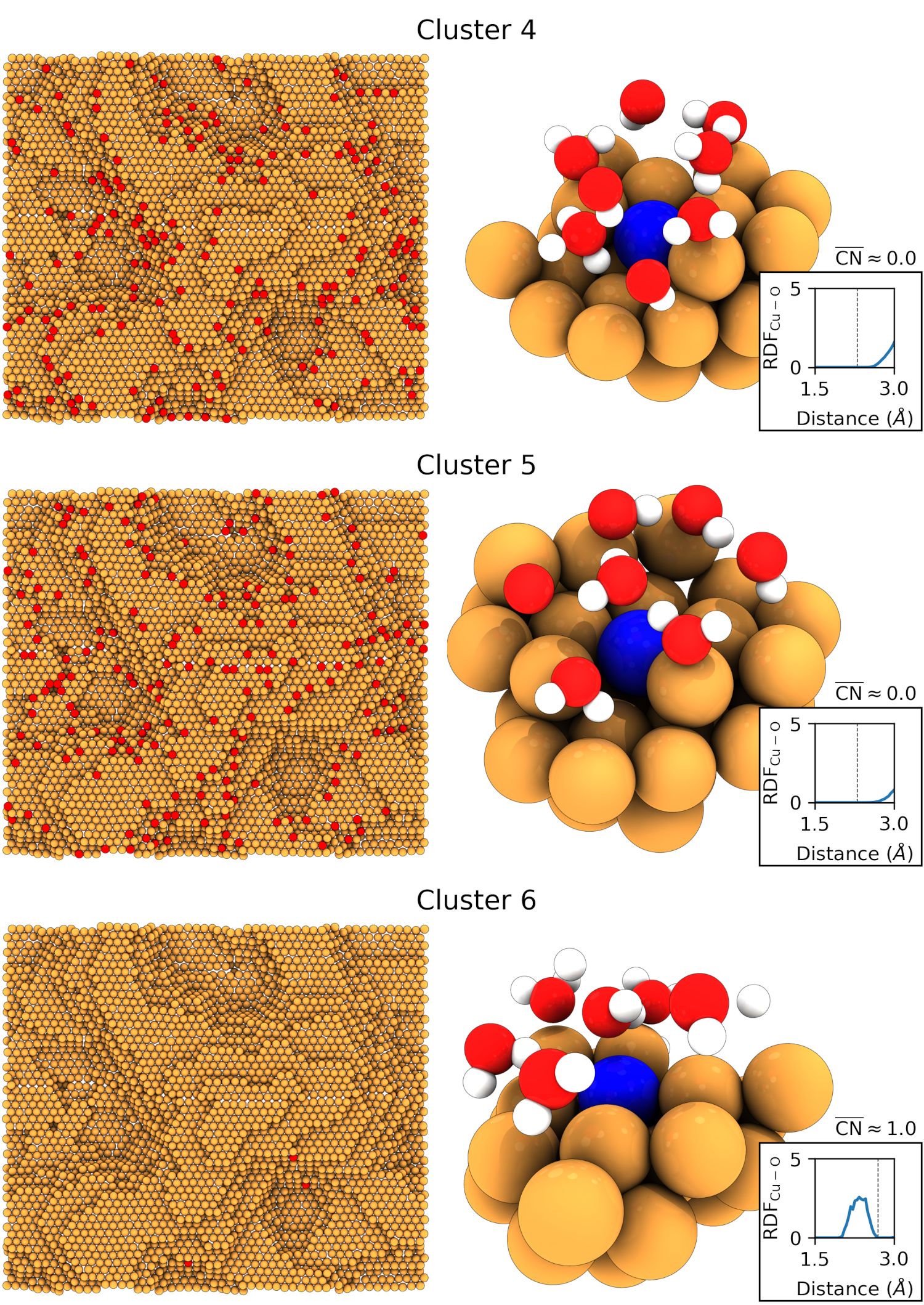}
    \caption{\textbf{Rough copper surface with atoms marked according to their cluster classification.} The left side of the figure depicts the rough copper surface, with atoms colored according to their cluster assignment as specified in Figure 6. On the right, an exemplary environment within a \SI{5}{\angstrom} radius of a cluster atom (marked in blue) is shown. In this representation, copper atoms are colored brown, oxygen atoms are colored red, and hydrogen atoms are colored white. Additionally, the average radial distribution function between copper atoms in this cluster and oxygen atoms is plotted. The average coordination number of copper with respect to oxygen is determined by integrating the first peak of the radial distribution function up to the point indicated by the black line.}
    \label{sfig:interface_clusters2}
\end{figure*}

\begin{figure*}
    \centering
    \includegraphics[width=0.8\textwidth]{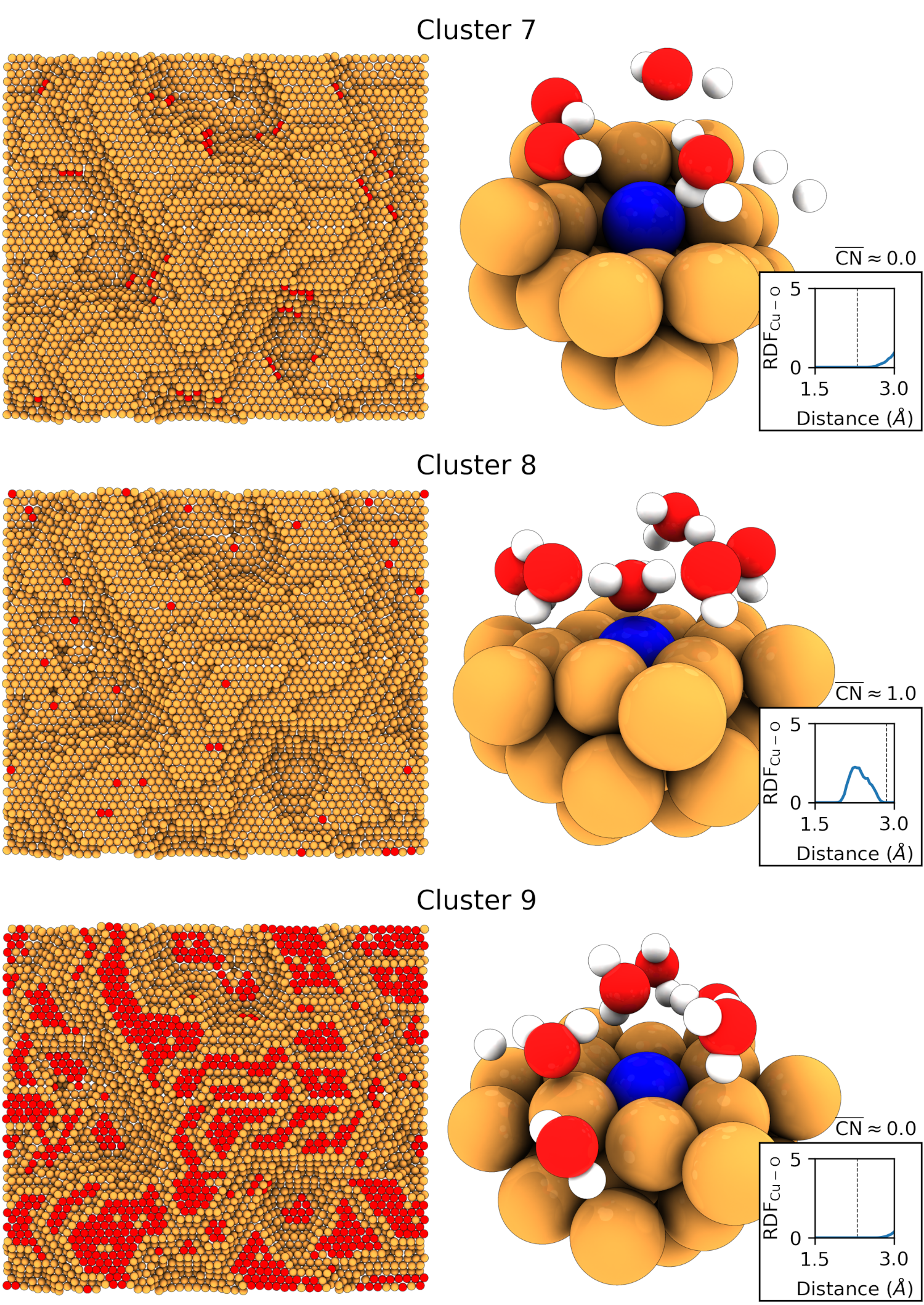}
    \caption{\textbf{Rough copper surface with atoms marked according to their cluster classification.} The left side of the figure depicts the rough copper surface, with atoms colored according to their cluster assignment as specified in Figure 6. On the right, an exemplary environment within a \SI{5}{\angstrom} radius of a cluster atom (marked in blue) is shown. In this representation, copper atoms are colored brown, oxygen atoms are colored red, and hydrogen atoms are colored white. Additionally, the average radial distribution function between copper atoms in this cluster and oxygen atoms is plotted. The average coordination number of copper with respect to oxygen is determined by integrating the first peak of the radial distribution function up to the point indicated by the black line.}
    \label{sfig:interface_clusters3}
\end{figure*}

\begin{figure*}
    \centering
    \includegraphics[width=0.8\textwidth]{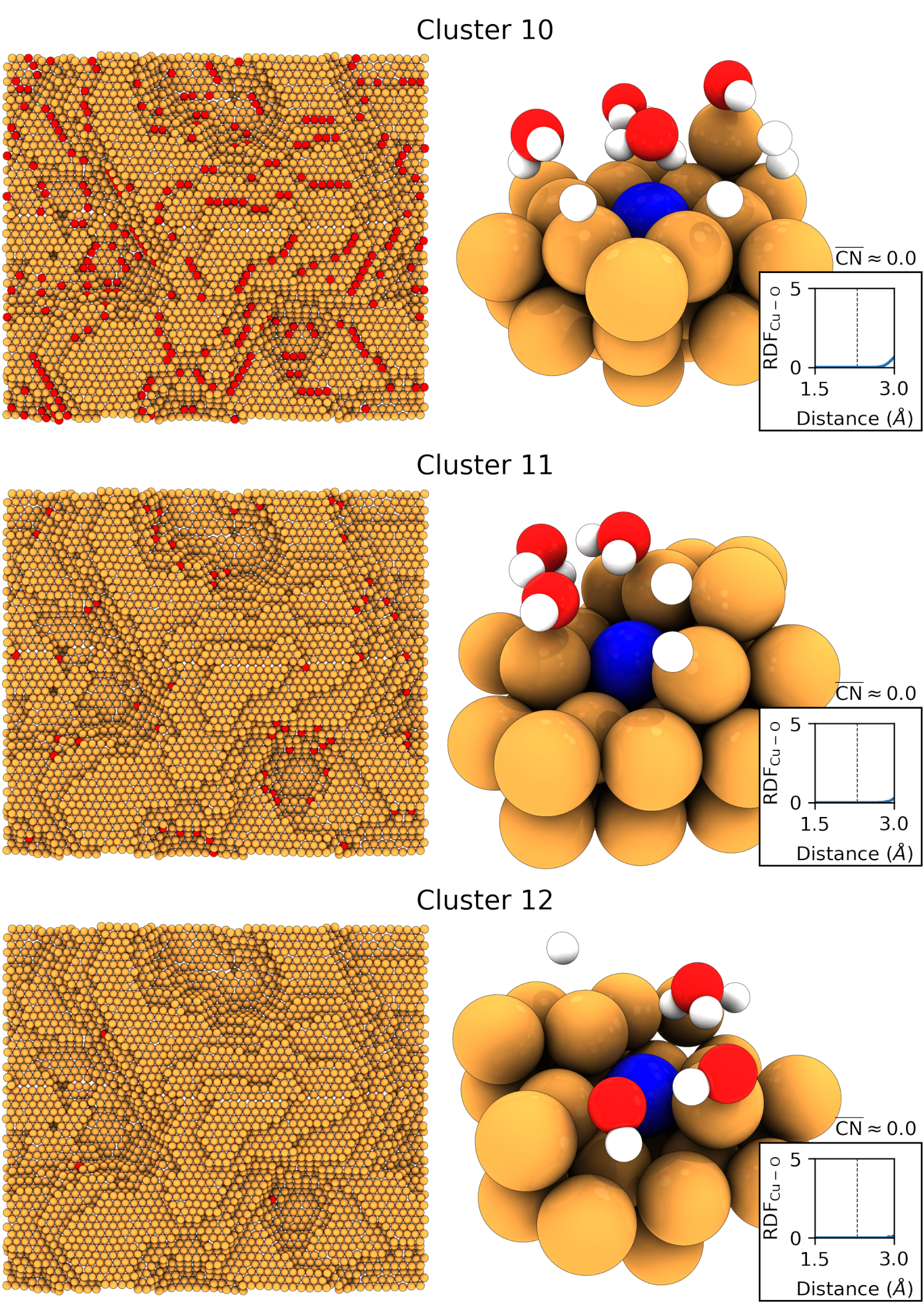}
    \caption{\textbf{Rough copper surface with atoms marked according to their cluster classification.} The left side of the figure depicts the rough copper surface, with atoms colored according to their cluster assignment as specified in Figure 6. On the right, an exemplary environment within a \SI{5}{\angstrom} radius of a cluster atom (marked in blue) is shown. In this representation, copper atoms are colored brown, oxygen atoms are colored red, and hydrogen atoms are colored white. Additionally, the average radial distribution function between copper atoms in this cluster and oxygen atoms is plotted. The average coordination number of copper with respect to oxygen is determined by integrating the first peak of the radial distribution function up to the point indicated by the black line.}
    \label{sfig:interface_clusters4}
\end{figure*}

\begin{figure*}
    \centering
    \includegraphics[width=0.8\textwidth]{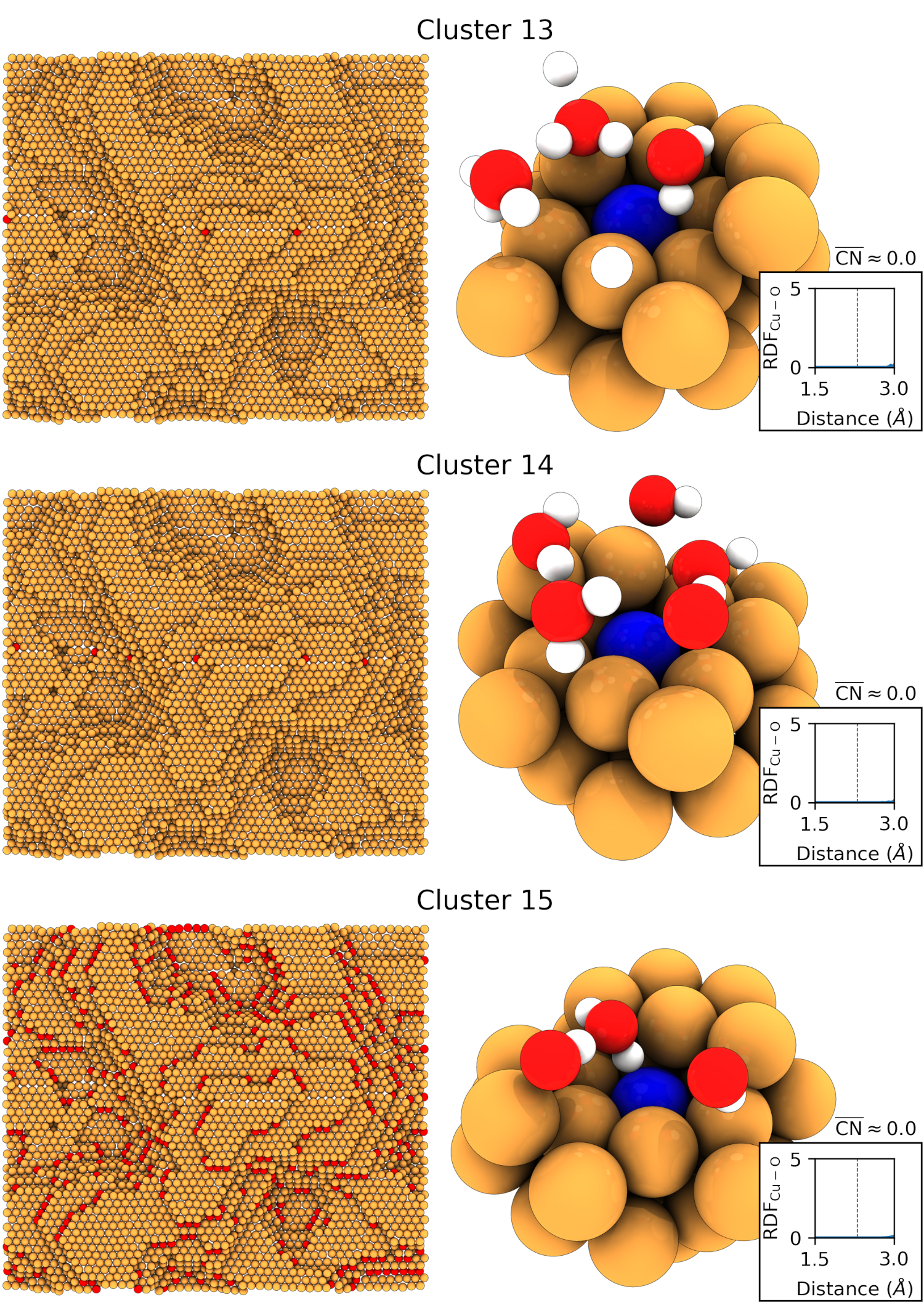}
    \caption{\textbf{Rough copper surface with atoms marked according to their cluster classification.} The left side of the figure depicts the rough copper surface, with atoms colored according to their cluster assignment as specified in Figure 6. On the right, an exemplary environment within a \SI{5}{\angstrom} radius of a cluster atom (marked in blue) is shown. In this representation, copper atoms are colored brown, oxygen atoms are colored red, and hydrogen atoms are colored white. Additionally, the average radial distribution function between copper atoms in this cluster and oxygen atoms is plotted. The average coordination number of copper with respect to oxygen is determined by integrating the first peak of the radial distribution function up to the point indicated by the black line.}
    \label{sfig:interface_clusters5}
\end{figure*}

\begin{figure*}
    \centering
    \includegraphics[width=0.8\textwidth]{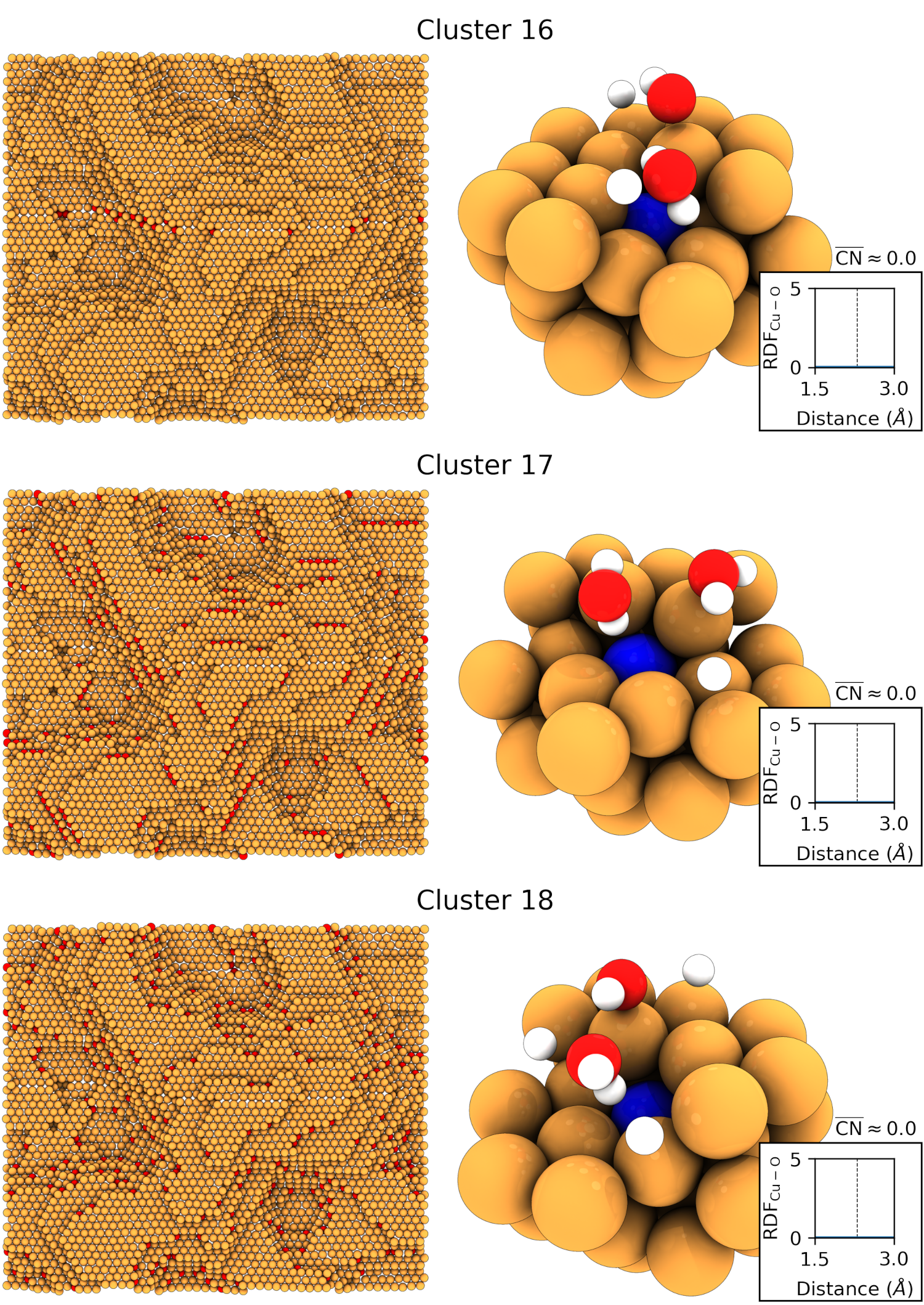}
    \caption{\textbf{Rough copper surface with atoms marked according to their cluster classification.} The left side of the figure depicts the rough copper surface, with atoms colored according to their cluster assignment as specified in Figure 6. On the right, an exemplary environment within a \SI{5}{\angstrom} radius of a cluster atom (marked in blue) is shown. In this representation, copper atoms are colored brown, oxygen atoms are colored red, and hydrogen atoms are colored white. Additionally, the average radial distribution function between copper atoms in this cluster and oxygen atoms is plotted. The average coordination number of copper with respect to oxygen is determined by integrating the first peak of the radial distribution function up to the point indicated by the black line.}
    \label{sfig:interface_clusters6}
\end{figure*}

\begin{figure*}
    \centering
    \includegraphics[width=0.8\textwidth]{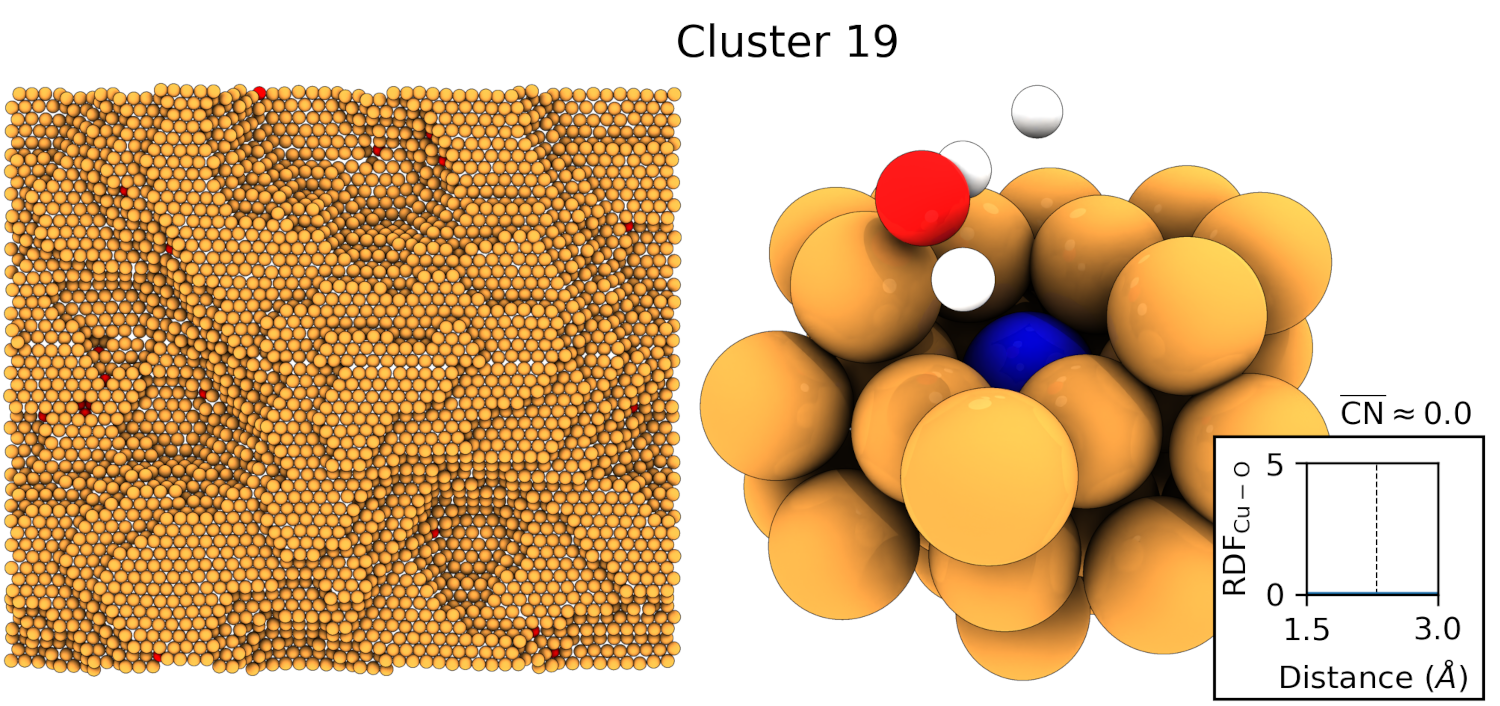}
    \caption{\textbf{Rough copper surface with atoms marked according to their cluster classification.} The left side of the figure depicts the rough copper surface, with atoms colored according to their cluster assignment as specified in Figure 6. On the right, an exemplary environment within a \SI{5}{\angstrom} radius of a cluster atom (marked in blue) is shown. In this representation, copper atoms are colored brown, oxygen atoms are colored red, and hydrogen atoms are colored white. Additionally, the average radial distribution function between copper atoms in this cluster and oxygen atoms is plotted. The average coordination number of copper with respect to oxygen is determined by integrating the first peak of the radial distribution function up to the point indicated by the black line.}
    \label{sfig:interface_clusters7}
\end{figure*}

\clearpage

\begin{table*}
    \caption{\textbf{Cluster classification.} Classification of all clusters from Figure 6 in the corresponding parts on the rough copper surface. For each cluster, we also note whether there is water chemisorbed.}
    \begin{tabular}{p{2cm}p{8cm}c}
         Cluster & Position & Water chemisorbed  \\
         \hline
         \hline
         \#1 & Corner of a (111) plane with two neighbors in the same plane. & Yes. \\
         \#2 & Corner of a (111) plane with three neighbors in the same plane. & Yes. \\
         \#3 & Edge of a (111) plane with four neighbors in the same plane. Partially atoms at the stacking fault. & Yes. \\
         \#4 & Edge of a (111) plane with four neighbors in the same plane. & No. \\
         \#5 & Edge of a (111) plane with five neighbors (nearly completely surrounded by copper atoms). & No. \\
         \#6 & (100) surface like facets. & Yes. \\
         \#7 & (100) surface like facets. & No. \\
         \#8  & (111) surface. & Yes. \\
         \#9  & (111) surface. & No. \\
         \#10 & (111) surface not directly below an edge, but close to. & No. \\
         \#11 & (100) surface like facets directly below edges. & No. \\
         \#12 & Rather complex edge/corner arrangements. & No. \\
         \#13 & Directly below edges at the stacking fault. & No. \\
         \#14 & Directly below edges at the stacking fault. & No. \\
         \#15 & Directly below edges at (111) surface facets. & No. \\
         \#16 & Closely below surface at stacking fault. & No. \\
         \#17 & Directly below edge of a (111) surface facet. & No. \\
         \#18 & Below (111) surface corners. & No. \\
         \#19 & In small holes nearly below the surface. & No. \\
    \end{tabular}
    \label{stab:classification_clusters}
\end{table*}